\documentclass[
superscriptaddress,
preprint,
8pt
amsmath,amssymb,
aps,
]{revtex4-1}
\usepackage{graphicx}
\usepackage{dcolumn}
\usepackage[usenames,dvipsnames,svgnames,table]{xcolor}
\usepackage[colorlinks=true,
linkcolor=blue,
urlcolor=black,
citecolor=blue]{hyperref}
\usepackage{bm}
\usepackage{enumerate}
\usepackage{lipsum}
\usepackage{epstopdf}
\usepackage{float}
\usepackage[caption = false]{subfig}
\usepackage{tabularx}
\usepackage{color}
\usepackage{longtable}
\usepackage{rotating}
\usepackage[utf8]{inputenc}

{\tiny {\tiny }}
\begin{document}

\preprint{APS/123-QED}

\title{Precise many-body calculations and hyperfine interaction effect on dynamic polarizabilities at the low-lying energy levels of Y$^{2+}$}

\author{Arghya Das}
\email{arghyadas@iitkgp.ac.in}
\affiliation{Department of Physics, Indian Institute of Technology Kharagpur, Kharagpur-721302, India.}
\author{Anal Bhowmik}
\email{abhowmik@campus.haifa.ac.il}
\affiliation{Haifa Research Center for Theoretical Physics and Astrophysics, University of Haifa, Haifa 3498838, Israel}
\affiliation{Department of Mathematics, University of Haifa, Haifa 3498838, Israel}
\author{Narendra Nath Dutta}
\affiliation{Department of Chemical Sciences, Indian Institute of Science Education and Research Mohali, Punjab-140306, India.}
\author{Sonjoy Majumder}
\email{sonjoym@phy.iitkgp.ernet.in}
\affiliation{Department of Physics, Indian Institute of Technology Kharagpur, Kharagpur-721302, India.}

\date{\today}

\begin{abstract}
The present work determines the precise values of magic wavelengths corresponding to the clock transitions 5$^2S$-4$^2D$ of Y$^{2+}$ ion both at the levels of fine-  and hyperfine-structures due to the external light beams having linear as well as circular polarization. To calculate the dynamic polarizabilities of the associated states of the transitions, we employ the sum-over-states technique, where the dominating and correlation sensitive part of the sum is evaluated using a highly correlated relativistic coupled-cluster theory. The estimated magic wavelengths of the light beams have substantial importance to cool and trap the ion using a blue-detuned trapping scheme. We  also present the  tune-out wavelengths which are useful in state-insensitive trapping and cooling. The vector component of a total polarizability, which is induced by a circularly polarized light only, can provide additional magic wavelengths. Considerable effects of hyperfine interaction on the values of polarizabilities and number of magic wavelengths divulge the importance of precise estimations of  hyperfine structure splitting.

\end{abstract}

\maketitle

\section{INTRODUCTION}

Accurate information about the dynamic polarizabilities of the atomic states at the hyperfine levels can be significant for quantum experiments such as trapping and cooling \cite{Kimble2003, Saf2013, Ferdinand2015}, atom interferometry \cite{Holmgren2012,Leonard2015},  quantum registers \cite{Gorshkov2009}, etc.. Also, appropriate environment for cooling and trapping of ions are absolutely necessary to perform error-free experiments for frequency standard \cite{Flambaum2008}, fundamental constants \cite{Kozlov2018,Porsev2009,Dutta2014}, quantum computer \cite{Haffner2008} and many other modern advanced technologies \cite{Lekitsch2015, Biedermann2015,Charriere2012,Dubetsky2006}. For instance, the preciseness of trapping and cooling parameters of ions mostly decides the fractional uncertainties of frequency standards which is sought to be of the order of $10^{-18}$ or less than this  \cite{Sahoo2018, Chou2010, Dube2014}. It has been shown over the past two decades that singly  \cite{Huntemann2016}, doubly \cite{Kozlov2013,Dutta2015}, or multiply \cite{Kozlov2018} charged ions can be competing candidates for the frequency standard in terms of accuracy and stability. Since the perturbation due to the external field reduces with increasing ionization for an atomic system \cite{Derevianko2012}, doubly ionized systems can be better choices  than singly ionized atoms for many of the above mentioned experiments. Besides the increasing ionicity, the other way to improve the accuracy is to prepare the states and transitions among them insensitive to external field  \cite{Dammalapati2016, Brown2017, Katori2009, Liu2015, Kaur2015, Roy2017, Bhowmik2018, Dutta2015}. 

The narrow line-width quadrupole transitions between ground states and long-lived metastable states of some moderate to heavy ions are being targeted from a long past for quantum experiments, such as atomic clock \cite{Tamm2014, Ludlow2015, Margolis2004, Oskey2006, Barwood2014}, quantum computer \cite{Bruzewicz2019}, etc.. Rubidium-like Sr$^{+}$, which has the ground state 5$^2S_{\frac{1}{2}}$, is a well-known ion for quantum technology and 5$^2S_{\frac{1}{2}}$ - 4$^2D_{\frac{3}{2},\frac{5}{2}}$ quadrupole transitions of this ion are  particularly utilized  in quantum computing \cite{Brownnutt2007, Eltony2016}. Here we are proposing the same narrow line-width transition of Y$^{2+}$, but now the ground state is 4$^2D_{\frac{3}{2}}$ instead of 5$^2S_{\frac{1}{2}}$. However, as indicated earlier, the advantage of using Y$^{2+}$ instead of Sr$^+$ is that the atomic states of Y$^{2+}$ are affected less by the perturbation of the electric field of the laser beam due to one unit more positive charge. But certainly, the effect of this perturbation can not be avoided for Y$^{2+}$ in general experimental circumstances.  Nevertheless, with the advent of cryogenic methods, highly charged ions are also possible to keep cool for a longer period \cite{Schwarz2012}, and therefore, Y$^{2+}$ can be extremely important for trap-assisted experimental studies.

\begin{figure}[htb]
\begin{center}
\includegraphics[width=0.8\textwidth]{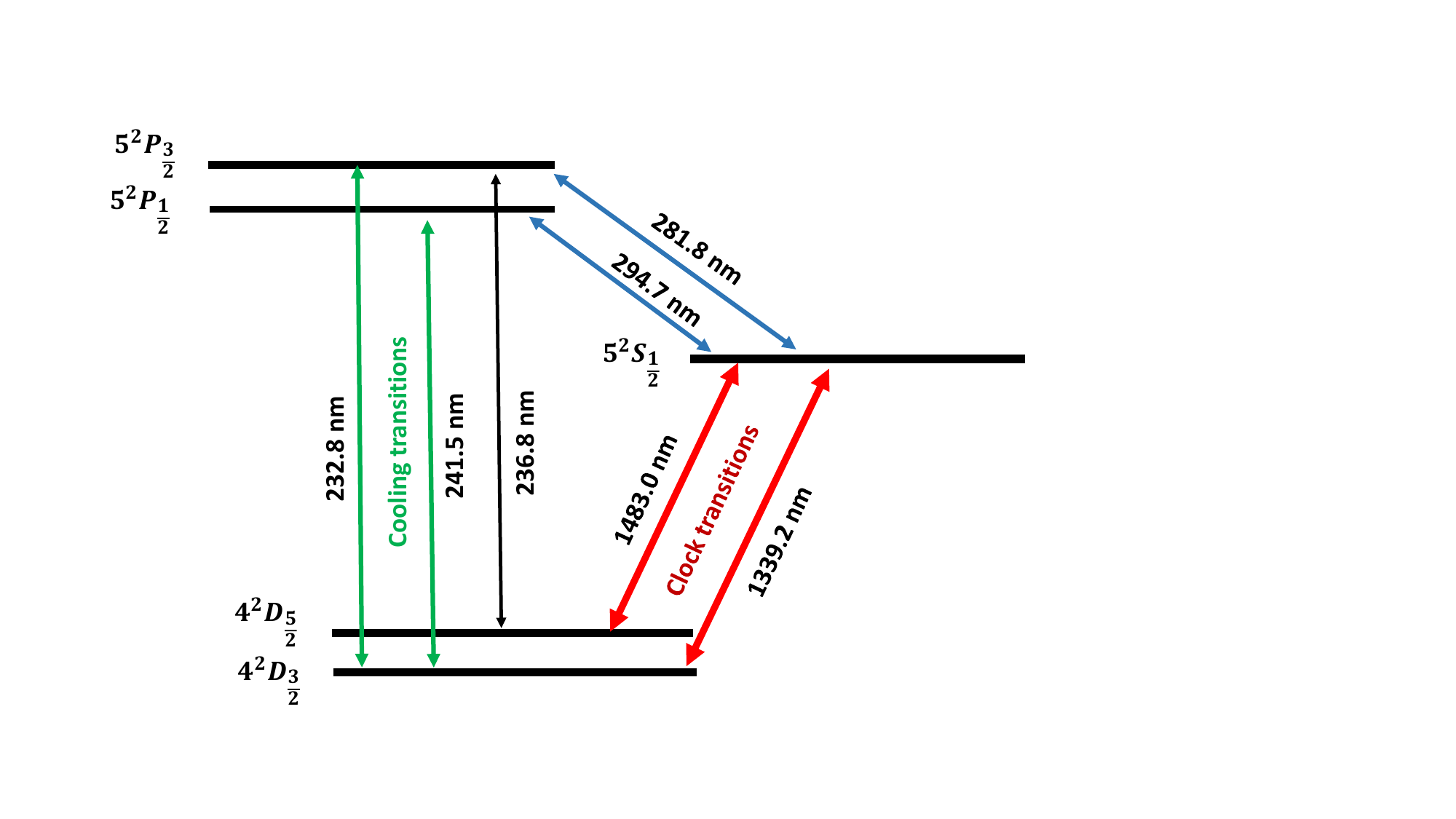}
\caption{(Color online) The clock (indicated by red color) and cooling (indicated by green color) transitions of Y$^{2+}$.}
\label{fig111}
\end{center}
\end{figure}

Further, a recent experiment on the isotope shift of $5 ^2S_{\frac{1}{2}}\rightarrow 5 ^2P_{\frac{1}{2}}$ transition of Y$^{2+}$ ion at 294.6 nm \cite{Vormawah2018}, where optical pumping $(4 ^2D_{\frac{3}{2}}\rightarrow 5 ^2P_{\frac{3}{2}})$ is achieved by 232.7 nm laser source, motivates us to propose $4 ^2D_{\frac{3}{2}} \rightarrow 5 ^2S_{\frac{1}{2}}$ transition (clock transition) for the above mentioned quantum experiments. This is evident from the energy-level diagram in Fig.~\ref{fig111} indicating possible cooling and clock transitions of Y$^{2+}$ among the level positions of all the five states. These prodigious technological and conceptual advancements on quantum experiments can not be worthwhile for precision measurements without the proper choice of wavelengths of the external field at which the differential ac-Stark shift between the associated energy levels of the atom or ion vanishes. The corresponding wavelengths are called magic wavelengths \cite{Mitroy2010}. In addition, the precise measurements of the tune-out wavelengths of light, where the polarizability of an atomic state vanishes, are crucial for atom interferometry \cite{Trubko2015}.

Whereas a linearly polarized beam can induce scalar and tensor components of a valence polarizability, the circularly polarized light can add a vector component to that. This vector part of the polarizability arises from the induced dipole moment perpendicular to the polarization of the field. As a consequence, use of a circularly polarized light is advantageous in many situations of the quantum experiments \cite{Flambaum2008}. The urge to find these magic and tune-out wavelengths inspires the theorists for highly accurate calculations on the dynamic polarizabilities of the electronic states of atoms or ions involved in the experiments. Here we are interested in evaluating dynamic polarizabilities of the ground (4$^2D_{\frac{3}{2}}$), first-excited (4$^2D_{\frac{5}{2}}$) and second-excited (5$^2S_{\frac{1}{2}}$) states of Y$^{2+}$. Another advantage of using Y$^{2+}$ is the relatively long lifetime of the first (244.08s) and second (10.76s) excited states \cite{Das2018} of this ion. In addition to these three clock states, the 5$^2P_{\frac{1}{2},\frac{3}{2}}$ states of Y$^{2+}$ are also important for the cooling transitions 4$^2D_{\frac{3}{2},\frac{5}{2}}$ $\rightarrow$ 5$^2P_{\frac{1}{2},\frac{3}{2}}$ achievable at 230-245 nm \cite{Vormawah2018, Meguro1994, Kumagai2004, Basting2004, Sakuma2004, Uchimura1999, Lipson2007}. Therefore, the precise knowledge of static and dynamic polarizabilities of these five low-lying states of Y$^{2+}$ is important in conducting trap-assisted precision experiments on this ion. Moreover, it is important to note that the scalar component of the static polarizability provides the theoretical estimation of the black-body radiation shift \cite{Mitroy2010, Jiang2009, BBR1,BBR2}, one of the important inputs to determine the clock accuracy.

Precise knowledge of hyperfine-structure constants or hyperfine constants of atoms and ions can be an important requirement in many experiments associated with the dynamic polarizabilities of hyperfine multiplets of the clock states. Experiments on the hyperfine splittings are also known as one of the first applications of trapped ions \cite{Werth1995, Hucul2017}. Moreover, the hyperfine constants are also important in an investigation of chemical composition of the Sun and stars. Being one of the astrophysically important elements \cite{Brage1998, Biemont2011, Maniak1994}, hyperfine constants of Y$^{2+}$ thus need a highly accurate theoretical treatment. So far, there has been one recent work \cite{Vormawah2018} along with a very old experiment \cite{Crawford1949} providing hyperfine constants only for 5$^2S_{\frac{1}{2}}$ and 5$^2P_{\frac{1}{2}}$ states of Y$^{2+}$. However, such values are also needed to be estimated for the other states as well theoretically and/or experimentally.  

We further calculate the dynamic polarizabilities of the 4$ ^2D_{\frac{3}{2},\frac{5}{2}}$ and 5$^2S_{\frac{1}{2}}$ states at the hyperfine levels and estimate the magic wavelengths for the 5$^2S_{\frac{1}{2}}$ $-$ 4$^2D_{\frac{3}{2}}$, 5$^2S_{\frac{1}{2}}$ $-$ 4$^2D_{\frac{5}{2}}$ clock transitions in the presence of a linearly and a circularly polarized light. For a particular hyperfine multiplet, the tune-out wavelength due to a circularly polarized light shifts with respect to the tune-out wavelength due to a linearly polarized light and this shift is a measure of fictitious magnetic field useful for trapping and cooling \cite{Kien2013}.  To calculate the dynamic polarizabilities of the clock states at both the fine- and hyperfine-structure levels and the static polarizabilities of the 5$^2P_{\frac{1}{2},\frac{3}{2}}$ states, we have employed a highly accurate relativistic many-body formalism. The detail of this formalism is discussed at the end of the Section II and the beginning of Section III.

\section{THEORY}
The Stark-shift $(\Delta\xi_v )$ for a single-valence atom or ion with the valence electron in the  `$v$'-th orbital is obtained by the second-order time-independent perturbation theory  \cite{Mitroy2010}
\begin{eqnarray}\label{1}
\Delta\xi_v(\omega)=\sum_{i \neq v}\frac{1}{\omega_{vi}}|\langle\psi_v|-\textbf{\textit{D}}\cdot \textbf{\textit{E}}|\psi_i\rangle|^2 =-\frac{1}{2}\alpha_v(\omega)\textit{E}^2
\end{eqnarray}
where $E$ is the strength of the applied electric field at the position of the atom, $-\textbf{\textit{D}}\cdot \textbf{\textit{E}}$ is the electric dipole interaction Hamiltonian, and  $\omega_{vi}=\epsilon_v-\epsilon_i$ is the  resonance frequency associated with the electric dipole ($E1$) transition between the states $|\psi_v\rangle$ and $|\psi_i\rangle$. Due to the frequency dependence of ${\textit{\textbf{E}}}$, i.e. $\textit{\textbf{E}}(\omega)$, the electric dipole polarizability $\alpha_v$ of the state $|\psi_v\rangle$  also becomes frequency-dependent $\alpha_v(\omega)$ and can be decomposed into three components \cite{Dutta2015,Mitroy2010}:
\begin{eqnarray}\label{2}
\alpha_v(\omega) = \alpha^C_v(\omega) + \alpha^{VC}_v(\omega) + \alpha^V_v(\omega) .
\end{eqnarray}
$\alpha^C_v(\omega)$  represents the contribution to the total polarizability due to the ionic core  \cite{Mitroy2010,Dutta2015}, and its value is irrespective of the presence of the valence electron at any orbital. It can be expressed as \cite{Mitroy2010, Dutta2015, Johnson1996} 
   \begin{eqnarray}\label{3}
  \alpha^C_v(\omega)= \frac{2}{3}\sum_{ap}\frac{|\langle\phi_a||d_{\text {DF}} ||\phi_p\rangle\langle\phi_a||d_{\text {RMBPT(2)}}||\phi_p\rangle|\omega_{pa}}{(\omega_{pa})^2 - \omega^2}.
  \end{eqnarray}
 The subscripts $a$ and $p$ in Eq.~(\ref{3}) indicate the core (fully occupied) and virtual (fully unoccupied) orbitals, respectively, with respect to the electron distributions for the ionic core. $\langle\phi_a||d_{\text {DF}}||\phi_p\rangle $ and $\langle\phi_a||d_{\text {RMBPT(2)}}||\phi_p\rangle$ represent the reduced electric dipole matrix elements obtained from the Dirac-Fock (DF) \cite{Grant2007, Motecc90} and the second-order relativistic many-body perturbation theory (RMBPT(2)) \cite{Johnson1996}, respectively. In one of the works regarding to Eq.~(\ref{3}), Mitroy \textit{et al.} \cite{Mitroy2010} mentioned an all-order theory like random phase approximation (RPA) \cite{Johnson1996} method with respect to our RMBPT(2) method. The conversion formula from a RPA matrix element to a corresponding RMBPT(2) matrix element is discussed in the work of Johnson \textit{et al.} \cite{Johnson1996} using diagrammatic perturbation theory. Nevertheless, for alkali-metal-like systems, the RMBPT(2) can provide a good approximation to the RPA in the calculation of core polarizability and this is tested by us for a number of such systems. $\alpha^{VC}_v(\omega)$ is the perturbation to the core polarizability in the presence of  the valence electron \cite{Dutta2015,Safronova2011}. This component of the polarizability is almost frequency independent within the frequency range considered in this work.

The estimation of $\alpha^V_v(\omega)$ is the most crucial part for the  calculation of the total dynamic polarizability and can be performed by using the following relationship \cite{Manakov1986}:
\begin{eqnarray}\label{4}
\alpha^V_v(\omega)= \alpha_v^{(0)}(\omega) + \sigma\frac{M_{J_v}}{2J_v} \alpha_v^{(1)}(\omega) + \frac{3M_{J_v}^2-J_v(J_v+1)}{J_v(2J_v-1)}\alpha_v^{(2)}(\omega).
\end{eqnarray}
Here $J_v$ and $M_{J_v}$ are the total angular momentum and its projection component for the state $|\psi_v\rangle$, respectively. The polarization factor $\sigma$ is zero for a linearly polarized light and $\pm 1$ for a circularly polarized light. This implies that the vector component of the valence polarizability is the only factor to differentiate between the impact caused by a linearly polarized light and the impact caused by a circularly polarized light. $\alpha_v^{(0)}(\omega)$,  $\alpha_v^{(1)}(\omega)$ and $\alpha_v^{(2)}(\omega)$ are the three components of the total valence polarizability specifying scalar, vector and tensor parts, respectively. Here, by `scalar', we only mean the scalar part of the valence polarizability, unless otherwise indicated. In the sum-over-states formalism, these components are expressed as \cite{Bhowmik2018,Manakov1986,Dutta2015} 
\begin{eqnarray}\label{5}
\alpha_{v}^{(0)}(\omega)= \frac{2}{3(2J_v + 1)}\sum_n d_{nv}
\end{eqnarray}
\begin{equation}\label{6}
\alpha_{v}^{(1)}(\omega)=-\sqrt{\frac{6J_v}{(J_v+1)(2J_v+1)}}\sum_n (-1)^{J_n+J_v} \left\{\begin{array}{ccc} J_v & 1 & J_v\\ 1 & J_n& 1 \end{array}\right \} 
\left(\frac{2\omega}{\omega_{nv}}\right)d_{nv},
\end{equation}
and
\begin{equation}\label{7}
\alpha_{v}^{(2)}(\omega)=4\sqrt{\frac{5J_v(2J_v-1)}{6(J_v+1)(2J_v+1)(2J_v+3)}}\sum_n (-1)^{J_n+J_v} \left\{\begin{array}{ccc} J_v & 1 & J_n\\ 1 & J_v& 2 \end{array}\right \} d_{nv},
\end{equation} 
where the apparently angular momentum independent factor 
$d_{nv}=
\{|\langle\psi_v||d||\psi_n\rangle|^2 \omega_{nv}\}/( \omega_{nv}^2-\omega^2)$
diverges at resonant frequency $\omega_{nv}=\epsilon_n-\epsilon_v$.
 
At a hyperfine energy level $F_v$ with nuclear spin $I$, the valence polarizability ($\alpha^{V}_{vF}(\omega)$) can be written in a similar mathematical form as of Eq.~(\ref{4}), but with replacement of total angular momentum and its projection quantum number ($J_v, M_{J_v}$) by the corresponding hyperfine and its projection quantum number ($F_v, M_{F_v}$) with \textit{$\textbf{F}_v =\textbf{J}_v+\textbf{I} $}. Also, the expression for the hyperfine-induced scalar part of the valence polarizability, $\alpha^{(0)}_{vF}(\omega)$, is equal to $\alpha_v^{(0)}(\omega)$ as the second-order scalar shift does not depend on any hyperfine quantum number \cite{Beloy2009}. However, the hyperfine-induced vector and tensor parts, $\alpha_{vF}^{(1)}(\omega)$ and $\alpha_{vF}^{(2)}(\omega)$, respectively, have extra factors following angular momentum algebra. These polarizabilities are related to the corresponding hyperfine-independent polarizabilities $\alpha^{(1)}_{v}(\omega)$ and $\alpha^{(2)}_{v}(\omega)$, respectively, by the following expressions \cite{Beloy2009,Dzuba2010,Kaur2017}:
\begin{equation}\label{8} 
 \alpha_{vF}^{(1)}(\omega)=(-1)^{J_v+F_v+I+1}\left\{\begin{array}{ccc} F_v & J_v & I\\ J_v & F_v& 1 \end{array}\right \}\sqrt{\frac{F_v(2F_v+1)(2J_v+1)(J_v+1)}{J_v(F_v+1)}}\alpha_v^{(1)}(\omega)
 \end{equation}
 and
 \begin{eqnarray}\label{9}
 \alpha_{vF}^{(2)}(\omega)=(-1)^{J_v+F_v+I}\left\{\begin{array}{ccc} F_v & J_v & I\\ J_v & F_v& 2 \end{array}\right \}&& \sqrt{\left(\frac{F_v(2F_v-1)(2F_v+1)}{(2F_v+3)(F_v+1)}\right)} \times \nonumber\\ 
&&\sqrt{ \left(\frac{(2J_v+3)(2J_v+1)(J_v+1)}{J_v(2J_v-1)}\right)}\alpha_v^{(2)}(\omega).
 \end{eqnarray}

The hyperfine shift (HFS) and splitting in the energy can be calculated accurately from the precise knowledge of hyperfine-structure constants corresponding to the magnetic dipole and electric quadrupole moments of the nucleus, which are known as the hyperfine $A$ and $B$ constants, respectively \cite{Bransden2006,Cheng1985}. The definitions of the above constants in terms of reduced matrix
elements  of the Hamiltonians corresponding to these nuclear moment interactions to the electronic sector are as follows: \cite{Dutta2013,Cheng1985}. 
 \begin{eqnarray}\label{10}
A&=&\mu_N g_I\frac{\langle
J_v||\textbf{T}^{(1)}||J_v\rangle}{\sqrt{J_v(J_v+1)(2J_v+1)}}
\end{eqnarray}
and
\begin{eqnarray}\label{11}
B&=&2eQ\sqrt{\frac{2J_v(2J_v-1)}{(2J_v+1)(2J_v+2)(2J_v+3)}}\langle
J_v||\textbf{T}^{(2)}||J_v\rangle,
\end{eqnarray}
where $\mu_N$ is the nuclear magneton, $g_I$ is the nuclear $g$-factor, and $Q$ is the quadrupole moment of the nucleus. The operators $\textbf{T}^{(1)}$ and $\textbf{T}^{(2)}$ can be written in terms of the spherical harmonic operators $Y_{kq}$ and the Dirac matrix $\bm{\alpha}$ using the following relations \cite{Cheng1985}:
\begin{equation}
\textbf{T}^{(1)}_q =\sum_j \textbf{t}^{(1)}_{q} = \sum_j -ie\sqrt{8\pi/3}r_j^{-2}\bm{\alpha}_j.\textbf{Y}_{kq}(\hat{r_j})
\end{equation} 
\begin{equation}
\textbf{T}^{(2)}_q=\sum_j \textbf{t}^{(2)}_q= \sum_j -er_j^{-3}\sqrt{\frac{4\pi}{2k+1}}\textbf{Y}_{kq}(\hat{r_j}) 
\end{equation} 
with $k=$ 1 and 2 for $\textbf{T}^{(1)}$ and $\textbf{T}^{(2)}$, respectively. The sum over $j$ implies the sum over all the electronic coordinates \cite{Cheng1985}. With the computed values of $A$ and $B$ constants, the HFS of an atomic energy level can be calculated almost precisely (the higher-order nuclear moments like nuclear octupole moment can contribute to HFS with a very little impact) using the formula \cite{Cheng1985}:
\begin{eqnarray}\label{14}
 E_{\text{HFS}} = \frac{AK}{2} + \frac{1}{2}\frac{3K(K+1)-4J_v(J_v+1)I(I+1)}{2I(2I-1)2J_v(2J_v-1)}B,
\end{eqnarray}
where $K=F_v(F_v+1)-I(I+1)-J_v(J_v+1)$.

The parameters which require exhaustive many-body treatment in the computations of the polarizabilities and hyperfine constants are the associated reduced matrix elements and the energy eigenvalues. In this work, we use three different theoretical approaches: DF, RMBPT(2) and relativistic coupled-cluster method with single, double and partial triple excitations (RCCSD(T)) \cite{Chaudhuri2003, Raghavachari1989}, in case-by-case basis with a compromise between accuracy and computational effort. This is elaborated in the next section. Nevertheless, very brief but adequate descriptions of the DF and RMBPT(2) theories are available in the work of Reiher and Hess \cite{Reiher2000}, and Johnson \textit{et al.}  \cite{Johnson1996}. The details of the DF theory with an algorithm of its computational implementation is given in \cite{Grant2007, Motecc90}, whereas the RMBPT(2) theory is discussed in \cite{Lindgren1986, Shavitt2009}. The RCCSD(T) is a well known many-body method for the calculations of electronic-structure properties accurately from a long time \cite{Bartlett2007, Shavitt2009, Lindgren1986, Lindgren1985, Lindgren1987, Dutta2016, Dixit2007, Bhowmik2017, Bhowmik2017b}. The RCCSD(T) method is discussed in detail in a review article of Bartlett and Musial \cite{Bartlett2007}; brief discussions on the different correlation mechanisms corresponding to the different level of cluster excitations and their numerical contributions can be found in our earlier works \cite{Dutta2016, Naren2013, Dutta2012, Bhowmik2017, Biswas2018}.

\section{RESULTS AND DISCUSSIONS} 

\subsection{Dynamic polarizabilities and magic wavelengths for fine-structure states}
We use sum-over-states formalism, as explained in the last section, to estimate the dynamic polarizabilities of  4$^2D_{\frac{3}{2},\frac{5}{2}}$ and 5$^2S_{\frac{1}{2}}$ states along with the static polarizabilities of these three states and 5$^2P_{\frac{1}{2},\frac{3}{2}}$ states. According to this formalism (see Eq. (\ref{5})-(\ref{7})), the accurate calculations of a large number of $E1$ matrix elements can be seen as essential for the accurate calculations of the scalar, vector and tensor parts of the total valence polarizability. The summation index $(n)$ in Eq. (\ref{5})-(\ref{7}) refers to the different intermediate single-valence open-shell states $|\psi_n\rangle$. These states correspond to different principal and angular momentum quantum numbers of the valence orbitals. The angular momentum quantum numbers are selected such that $\langle\psi_v||d||\psi_n\rangle \neq 0$ according to the $E1$ selection rule. In the present calculations, the  principal quantum number is considered up to 25, beyond which the $E1$ matrix elements have very negligible contributions to the total valence polarizabilities. 

As mentioned in the last paragraph of the previous section, depending on the comparative strengths of the $E1$ matrix elements in evaluating the valence polarizabilities, we consider three different levels of many-body theories: RCCSD(T), RMBPT(2) and DF. However, the former two correlated many-body theories are also based on the generations of DF orbitals to construct the zeroth-order wavefunctions. Nonetheless,  these DF orbitals are constructed using Gaussian-type-orbital (GTO) basis functions \cite{Motecc90, Dutta2016} of the type $r^{n_\kappa}e^{-\alpha_i r^2}$ with $\alpha_i=\alpha_0\beta^{i-1}$, where $\alpha_0$ and $\beta$ are considered as 0.00525 and 2.73, respectively. The values of $i$ are 1, 2, 3,...$N$, where $N$ is the size of the basis set \cite{Motecc90} for each symmetry. $n_\kappa$ is a constant whose values are 1, 2, 3, 4, 5, and 6 for $s$, $p$, $d$, $f$, $g$, and $h$-symmetries, respectively. \cite{Motecc90}. The size of the basis set considered for the abovementioned symmetries are 33, 30, 28, 25, 21, and 20, respectively \cite{Das2018}. The RCCSD(T) and RMBPT(2) calculations also require adequate inclusions of active-orbital basis sets to satisfy the correlation energy convergence criteria \cite{Dutta2016}, which are considered 11, 11, 13, 11, 10, and 6 in numbers for the $s$, $p$, $d$, $f$, $g$, and $h$-symmetries, respectively. 
 
In the sum-over-states strategy, the most dominant contribution to a valence polarizability appears from the sum (Eq.~(\ref{5})-(\ref{7})) of the terms having $E$1 matrix elements associated with the first few low-lying states. Accordingly, the excited states 5-8$^2S_{\frac{1}{2}}$, 5-8$^2P_{\frac{1}{2},\frac{3}{2}}$, 4-7$^2D_{\frac{3}{2},\frac{5}{2}}$ and 4-6$^2F_{\frac{5}{2},\frac{7}{2}}$ are found to be the most important states in the present sum-over-states formalism. The matrix elements are calculated here using the correlation exhaustive RCCSD(T) method, and the corresponding transition energies to calculate valence polarizability are extracted from the website of National Institute of Standards and Technology (NIST) \cite{nist}. Relatively less important  $E1$-matrix elements of the sum  are calculated by using RMBPT(2) \cite{Johnson1996} which includes core polarization corrections on top of the DF contributions \cite{Blundell1992,Lindgren1986}. This part sums up the contributions from the next five single valence states for all the symmetries. The $E1$ matrix elements, whose contributions have little significance to the total valence polarizabilities, are associated with the single valence states 14-25$^2S_{\frac{1}{2}}$, 14-25$^2P_{\frac{1}{2},\frac{3}{2}}$, 13-25$^2D_{\frac{3}{2},\frac{5}{2}}$ and 12-25$^2F_{\frac{5}{2},\frac{7}{2}}$. The wavefunctions and the corresponding $E1$ matrix elements of these states are calculated using the DF method only.

In Table \ref{table1}, we represent total static polarizabilities ($\alpha_v(0)$) of 4$^2D_{\frac{3}{2},\frac{5}{2}}$, 5$^2S_{\frac{1}{2}}$ and 5$^2P_{\frac{1}{2},\frac{3}{2}}$ states for the case $J_v=M_{J_v}$. The contributions from $\alpha^C_v(0)$ and $\alpha_v^{VC}(0)$ to the total static polarizabilities are also presented along with the scalar ($\alpha^{(0)}_v(0)$) and tensor ($\alpha^{(2)}_v(0)$) components of $\alpha^V_v(0)$. The decomposition of $\alpha^V_{v}(0)$ in terms of the RCCSD(T), RMBPT(2), and DF parts as discussed in the previous paragraph of this section is shown quantitatively in Table \ref{table11}. Nevertheless, our calculated values of the polarizabilities for the clock states in Table \ref{table1} are compared with the corresponding theoretical values of Safronova and Safronova \cite{Safronova2013}. They used an all-order relativistic many-body perturbation theory, which is very similar to a linearized coupled-cluster theory, for their calculation to the valence part of a polarizability. Our RCCSD(T) method, which is employed to calculate the most dominating part of a valence polarizability, is theoretically more accurate than their all-order method as the former is augmented with important non-linear terms. However, their method correspond to use of RPA matrix element with respect to our RMBPT(2) matrix element to compute the core polarizability \cite{Mitroy2010, Safronova2013, Johnson1983}. As mentioned in the previous section, our calculated value of core polarizability 4.28 a.u. is an approximation to the all-order RPA value of core polarizability 4.05 a.u.. Here it should be mentioned that indeed our strategy of calculating core polarizability is not highly accurate towards computing a total polarizability value. But this slight inaccuracy does not affect a calculated value of magic wavelength as the contribution of core polarizability to the total polarizabilities for the clock states is cancelled to determine this wavelength. However, this inaccuracy can affect a tune-out wavelength, which is not more than $\pm$1\AA$ $ for the calculated tune-out wavelengths discussed in the last paragraph of the present section.  The comparison in Table \ref{table1}  indicates 4.7\% deviation for 4$^2D_{\frac{3}{2}}$ state,  4.9\% deviation for 4$^2D_{\frac{5}{2}}$ state and 3.8\% deviation for 5$^2S_{\frac{1}{2}}$ state in the total static polarizability values. We also present the static polarizabilities for 5$^2P_{\frac{1}{2},\frac{3}{2}}$ states, which can be helpful in future experimental explorations of state-specific properties.

\begin{figure}[H]
\subfloat[]{\includegraphics[width=0.52\linewidth]{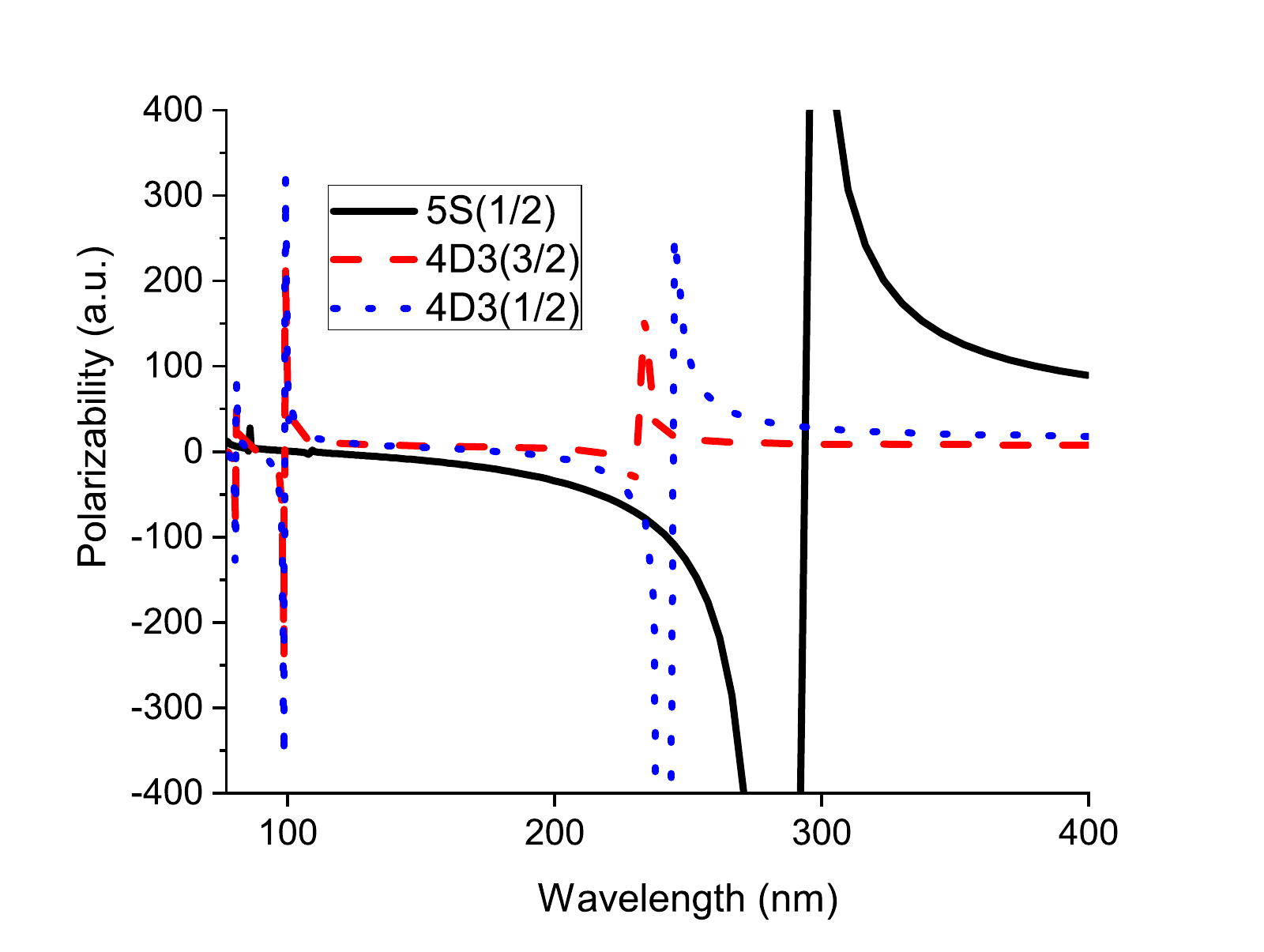}}
\subfloat[]{\includegraphics[width=0.52\linewidth]{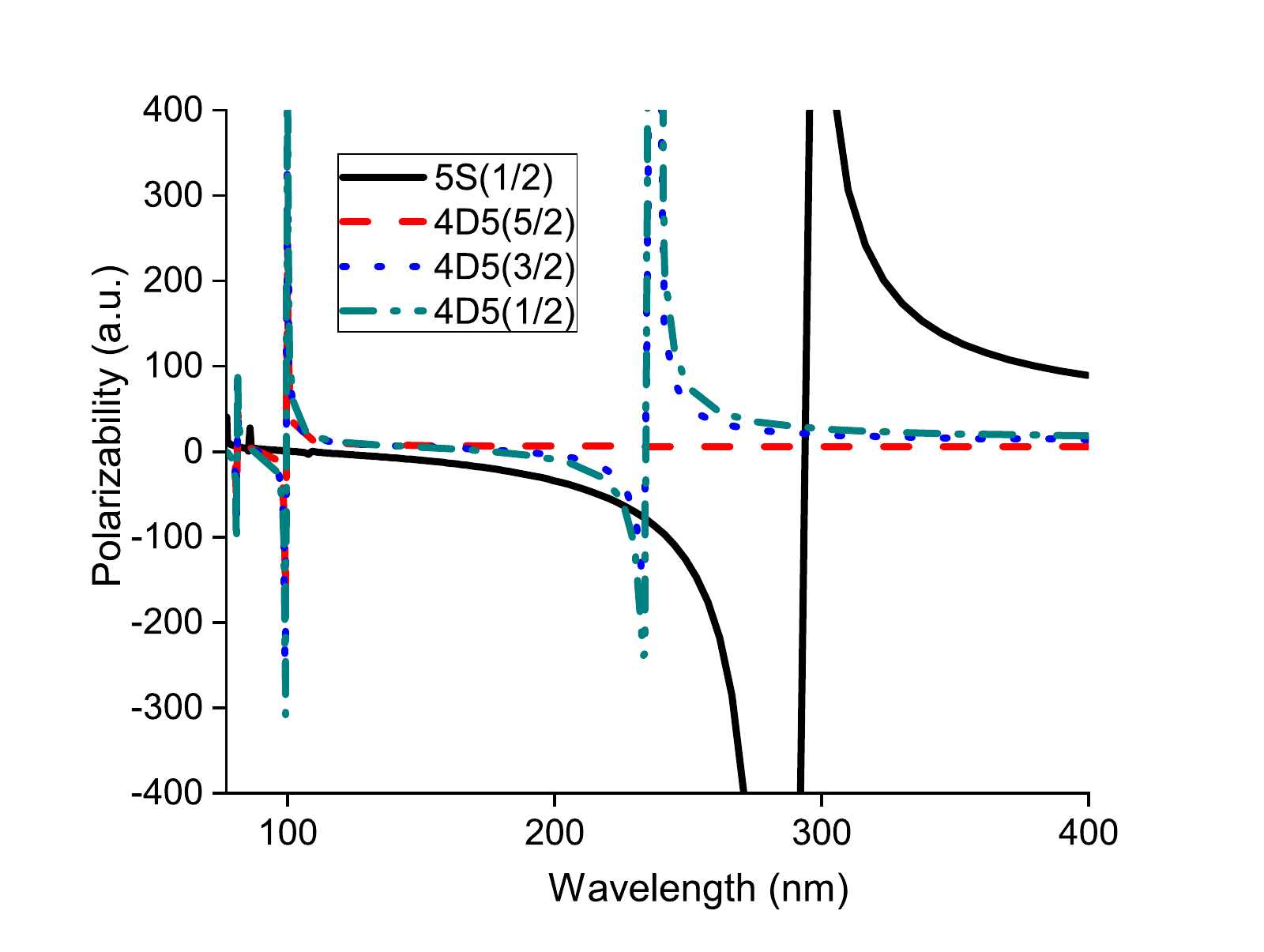}}\\
\subfloat[]{\includegraphics[width=0.52\linewidth]{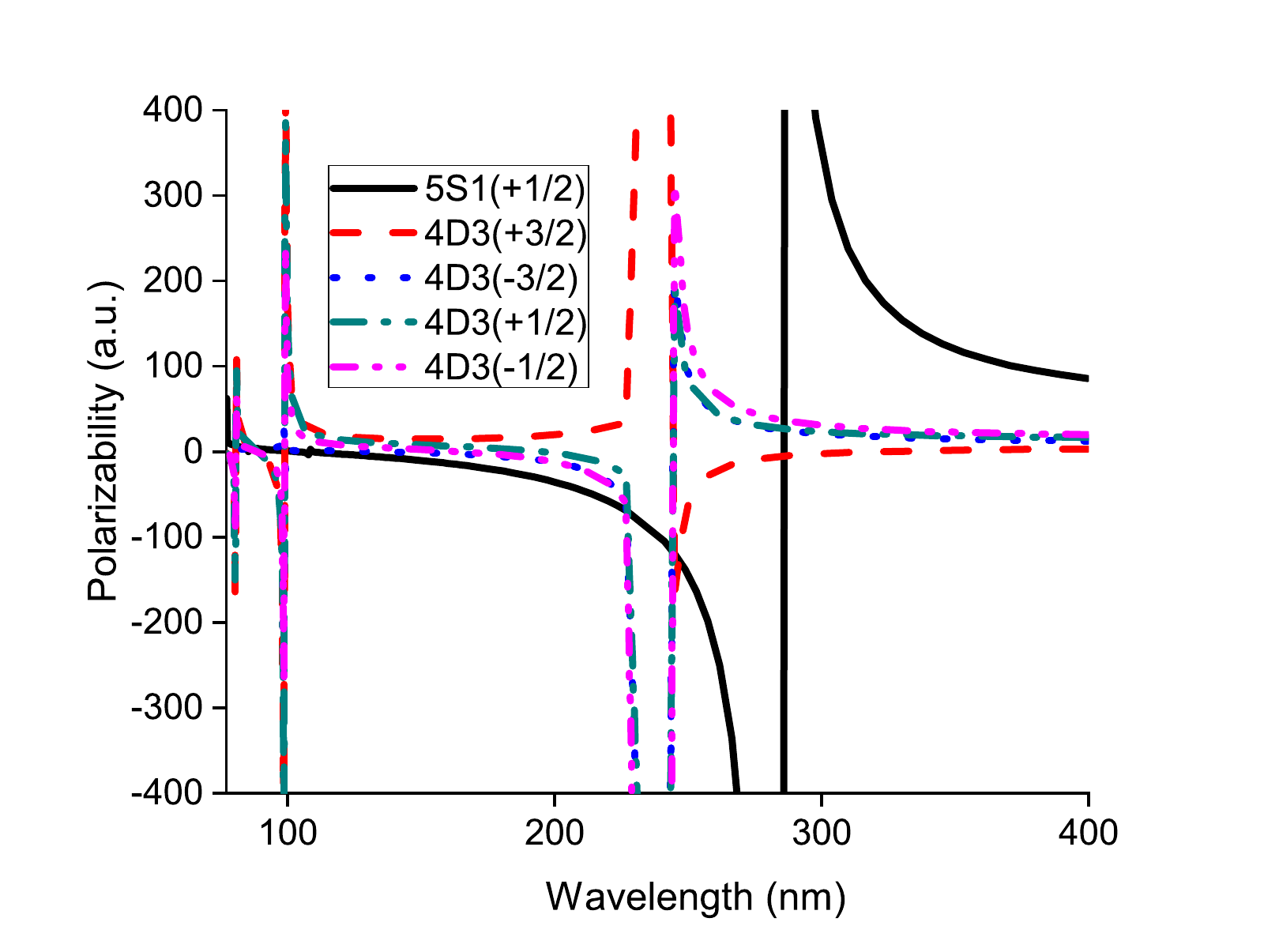}}
\subfloat[]{\includegraphics[width=0.52\linewidth]{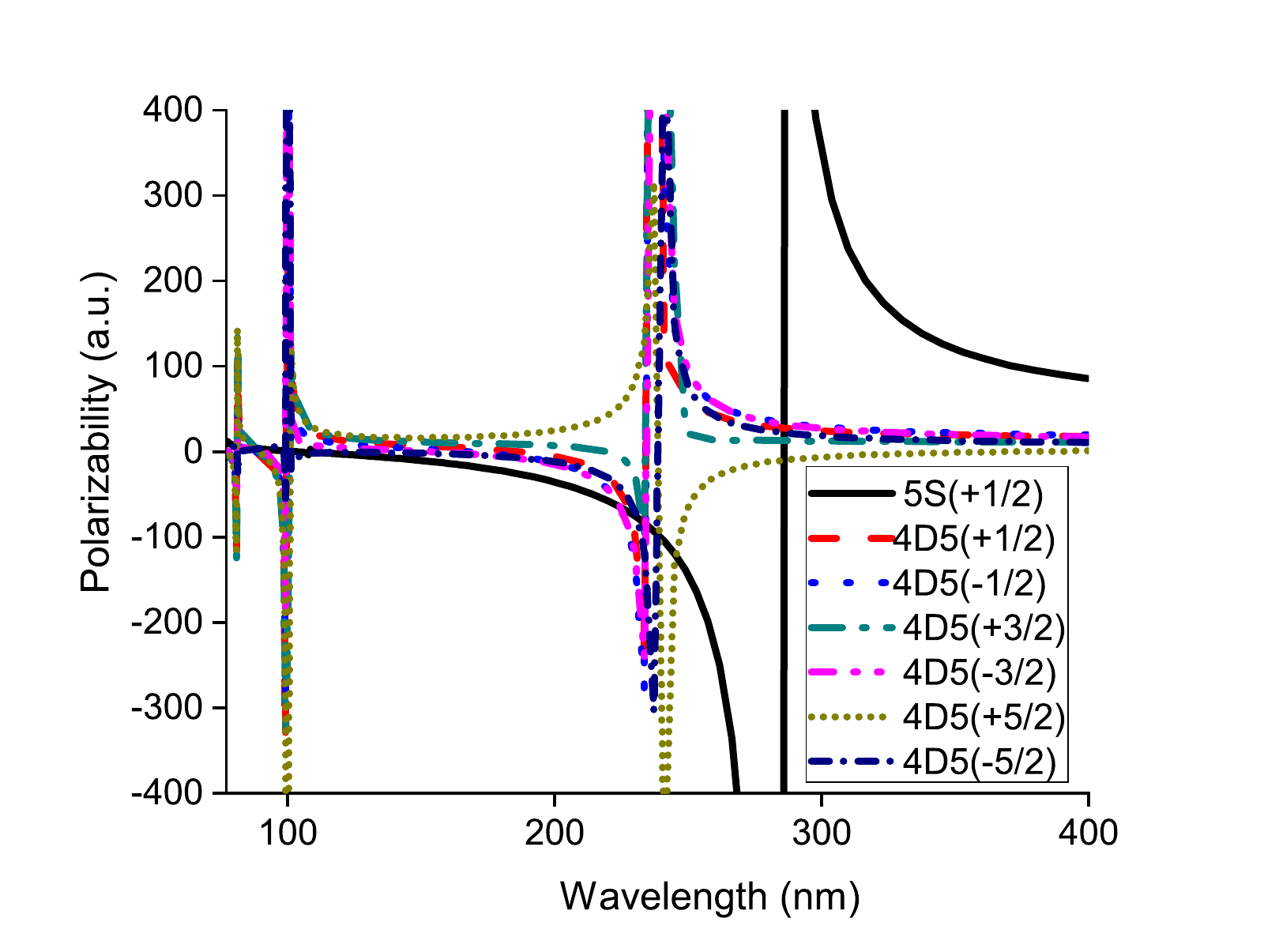}}
\caption{ (Color online) The variations of total polarizabilities (indicated by polarizability) with wavelengths for the states  5$^2S_\frac{1}{2}$ and 4$^2D_{\frac{3}{2}, \frac{5}{2}}$ to extract magic wavelengths for the transitions between them. (a) and (b) are for linearly polarized light.  (c) and (d) are for circularly polarized light. Here the different $(J_v, M_{J_v})$  levels of the state n$^{2}L_{J_v}$ are indicated by n$L2J_{v}(M_{J_v})$ with n is the principal quantum number of the valence orbital and $L$ is the orbital angular momentum quantum number of the ion.}
\label{fig1}
\end{figure}

Fig.~\ref{fig1} shows the wavelength dependency of the total polarizabilities for the 5$^2S_{\frac{1}{2}}$ and 4$^2D_{\frac{3}{2},\frac{5}{2}}$ states due to a linearly and a circularly polarized light. We consider right circularly polarized light (i.e., $\sigma=+1$ in Eq. (\ref{4})) in the present calculations of polarizabilities at the various $M_{J_v}$ levels of the clock states. Here we choose the most important region of the electromagnetic spectrum from $\lambda=77$ nm to $\lambda=400$ nm as there is no significant magic wavelength which can be prescribed for laser trapping purpose outside of this region. The entire spectrum of the wavelength as considered in all the four plots of Fig.~\ref{fig1} span from the vacuum ultraviolet (VUV) to the starting zone of the visible region. The peaks in the polarizability curves represent the resonances associated with the transitions 5$^2S_{\frac{1}{2}}$  $\rightarrow$ (5-7)$^2P_{\frac{1}{2},\frac{3}{2}}$ and  4$^2D_{\frac{3}{2},\frac{5}{2}} \rightarrow$ (5-7)$^2P_{\frac{1}{2},\frac{3}{2}},$ (4-5)$^2F_{\frac{5}{2},\frac{7}{2}}$. It is interesting to see that there is no resonance line in between 100 nm to 400 nm for 4D5(5/2) ($4^2D_{(J_v, M_{J_v})}$ state with $J_v=\frac{5}{2}$ and $M_{J_v}= \frac{5}{2}$) with a linearly polarized light (Fig.~\ref{1}(b)). This feature is a consequence of strong cancellation between scalar and tensor parts of the polarizability in that wavelength span. A similar characteristic for the resonance behaviour of 3D5(5/2) state was seen in the recent work on dynamic polarizability for Sc$^{2+}$ ion \cite{Dutta2015}. Indeed, the resonances are appeared for this state when one considers circular polarization of light. The circular polarization induces non-vanishing vector component which is the sole reason for the occurrences of resonances in the polarizability profile of 4D5(5/2).  
 
The crossing points between the curves representing the dynamic polarizabilities  of $5^2S_{\frac{1}{2}}$ and 4$^2D_{\frac{3}{2},\frac{5}{2}}$ states in Fig.~\ref{fig1} show a number of magic wavelengths for the transitions between them. We find quite a few magic wavelengths with high polarizability values at the mid-UV region (200-300 nm), and therefore, these wavelengths can be much significant for Y$^{2+}$ clock experiment with the best possible precision. All the magic wavelengths within the range of 77 nm to 400 nm are presented in Table~\ref{II} with the corresponding polarizability values. This table also reveals effect of the vector part of polarizability by comparing the results for the circularly polarized light with the results for a linearly polarized light. Here it is to be mentioned that the orientation of the polarization of light ($\sigma$) and the sign of $M_{J_v}$ decide the resultant sign of the vector part of polarizability (See Eq. (\ref{4})). This essentially means that the signs of these two factors ($\sigma$ and $M_{J_v}$) are responsible apart from the sign of $\alpha^{(1)}_v(\omega)$ to determine whether there should be an additive or a subtractive effect to the total polarizability for a linearly polarized light to achieve the total polarizability for the circularly polarized light. Nevertheless, due to the presence of the vector part in the polarizability, the circularly polarized light  provides relatively more number of magic wavelengths for the clock transitions and many of them are associated with large polarizability values. Moreover, both the values of the magic wavelengths and the corresponding polarizabilities are changed moderately in few cases due to the change in light polarization from linear and circular. This facilitates external control on slight tuning of the magic wavelengths. Since the wavelengths of the  transitions 5$^2S_{\frac{1}{2}}$ -- 4$^2D_{\frac{3}{2},\frac{5}{2}}$ are 1339.2 nm and 1483.0 nm, respectively,  all the magic wavelengths presented in the table support blue-detuned trapping scheme.

\subsection{Dynamic polarizabilities and magic wavelengths for hyperfine-structure states}  

Instead of considering the electronic fine-structure transitions, experimentalists prefer to consider hyperfine transitions in most cases of trapping and cooling processes. Therefore, it may be physically more meaningful to estimate the magic wavelengths for the transitions between different hyperfine levels of the clock states. 
 
The hyperfine-structure constant $A$ values for nine low-lying states of Y$^{2+}$ are calculated using the RCCSD(T) method and presented in Table~\ref{III}. In order to calculate these constants, we choose the most abundant isotope of Y with mass number $89$, nuclear spin ($I$) =$1/2$, and nuclear magnetic moment $(\mu)=-0.1374154$ $\mu_N$.  The nuclear charge distribution of this isotope is assumed to have the Fermi type form \cite{Visscher1997}. We compare the present RCCSD(T) values with the available hyperfine $A$ values in the literature and find good agreement between them \cite{Vormawah2018}. As the nuclear quadrupole moment is zero for a spin-half nucleus, the hyperfine shifts (see Eq.~(\ref{14})) and consequently the splittings are calculated using hyperfine $A$ constants only. The splitting values are displayed in the same table with comparison to a few very old experimental measurements \cite{Crawford1949}. This table also shows high values of relative correlation corrections to the hyperfine $A$ constants of the 4$^2D_{\frac{5}{2}}$ and 4$^2F_{\frac{5}{2}, \frac{7}{2}}$ states. This relative correlation correction is defined by $|\frac{\text{RCCSD(T) value}-\text{DF value}}{\text{DF value}}|\times 100$. From our investigation, we find strong core polarization effect \cite{Lindgren1985} is the main reason for such high impact of correlation. 
                                     
\begin{figure}[H]
\centering
\subfloat[]{\includegraphics[width=0.52\linewidth]{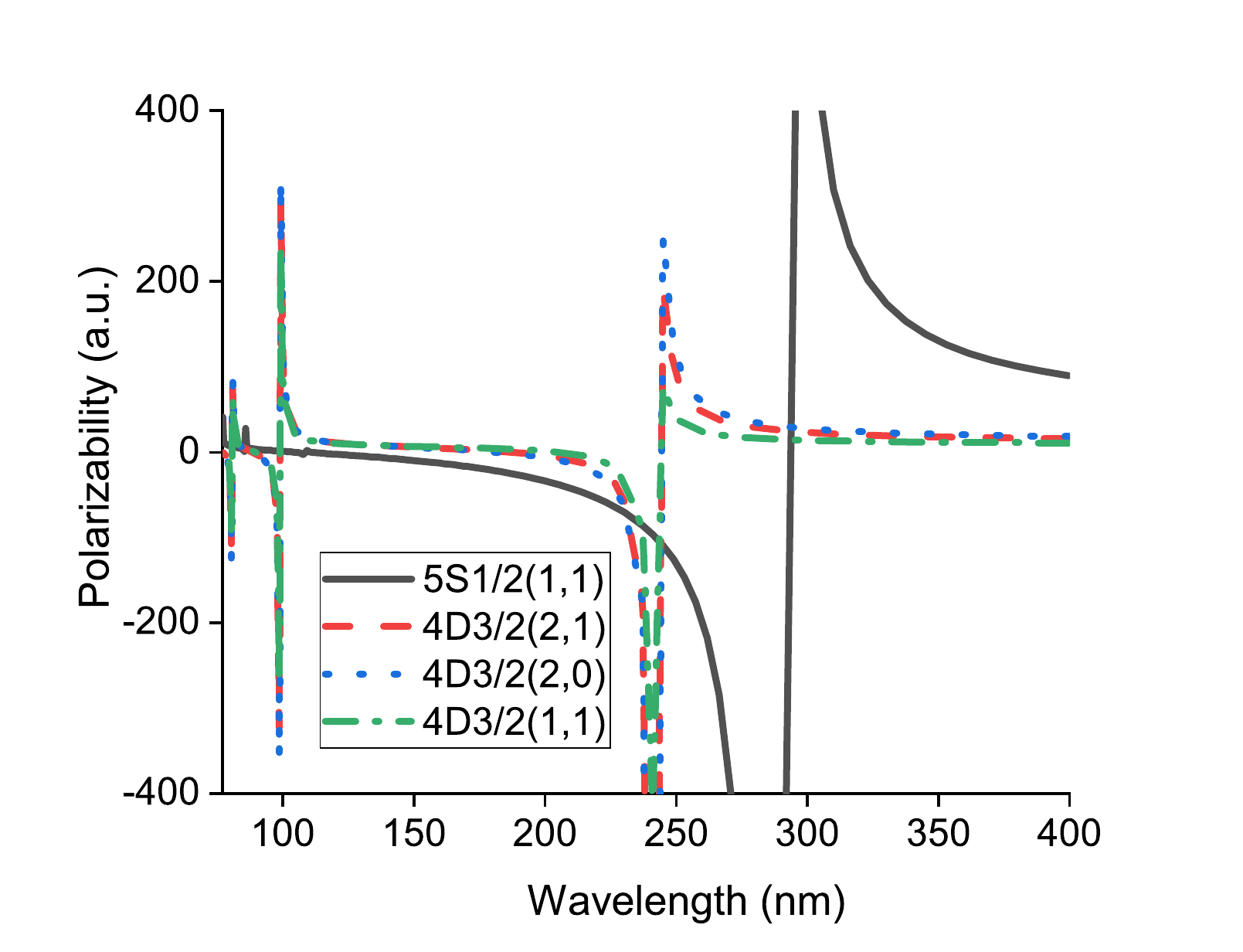}}
\subfloat[]{\includegraphics[width=0.52\linewidth]{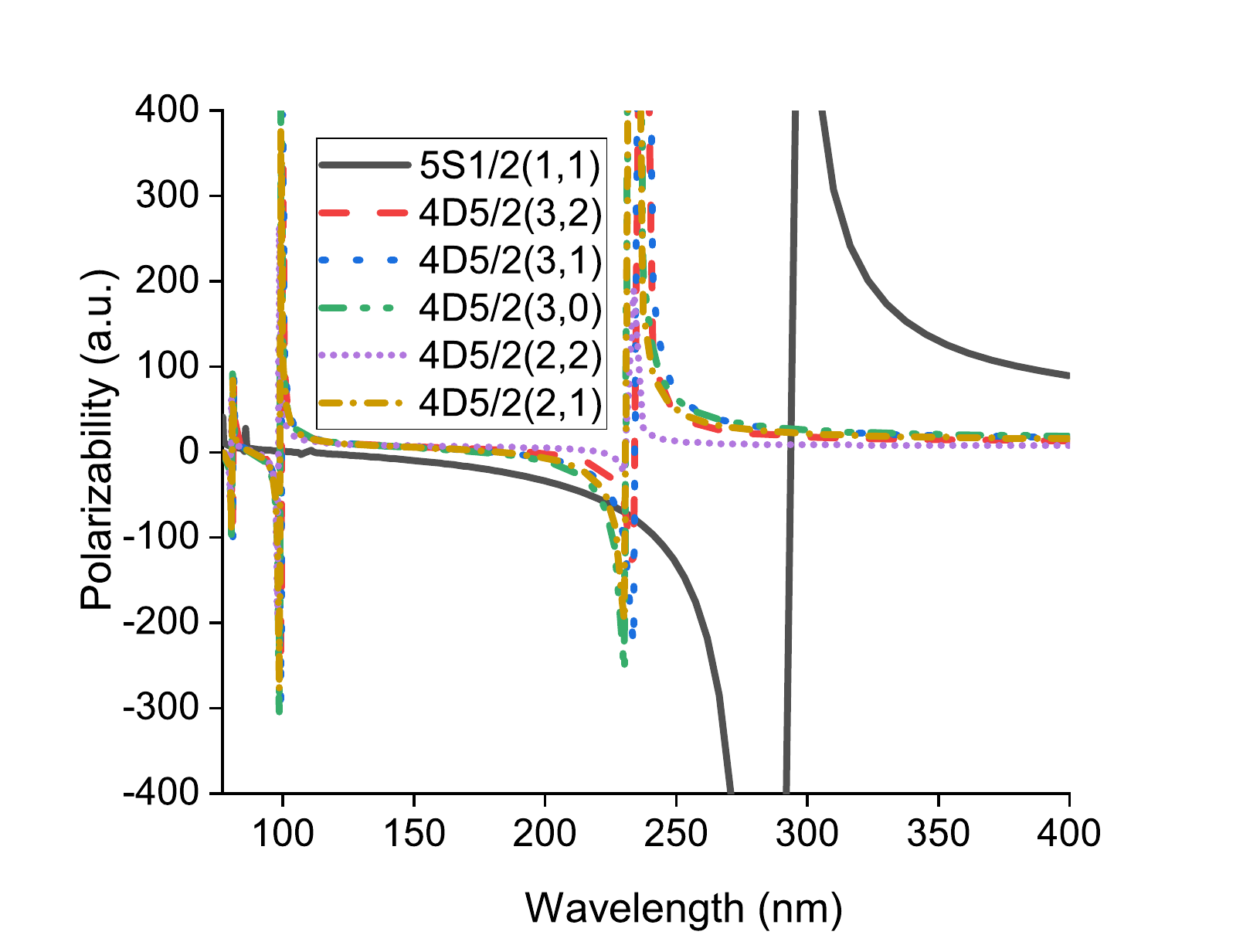}}\\
\subfloat[]{\includegraphics[width=0.52\linewidth]{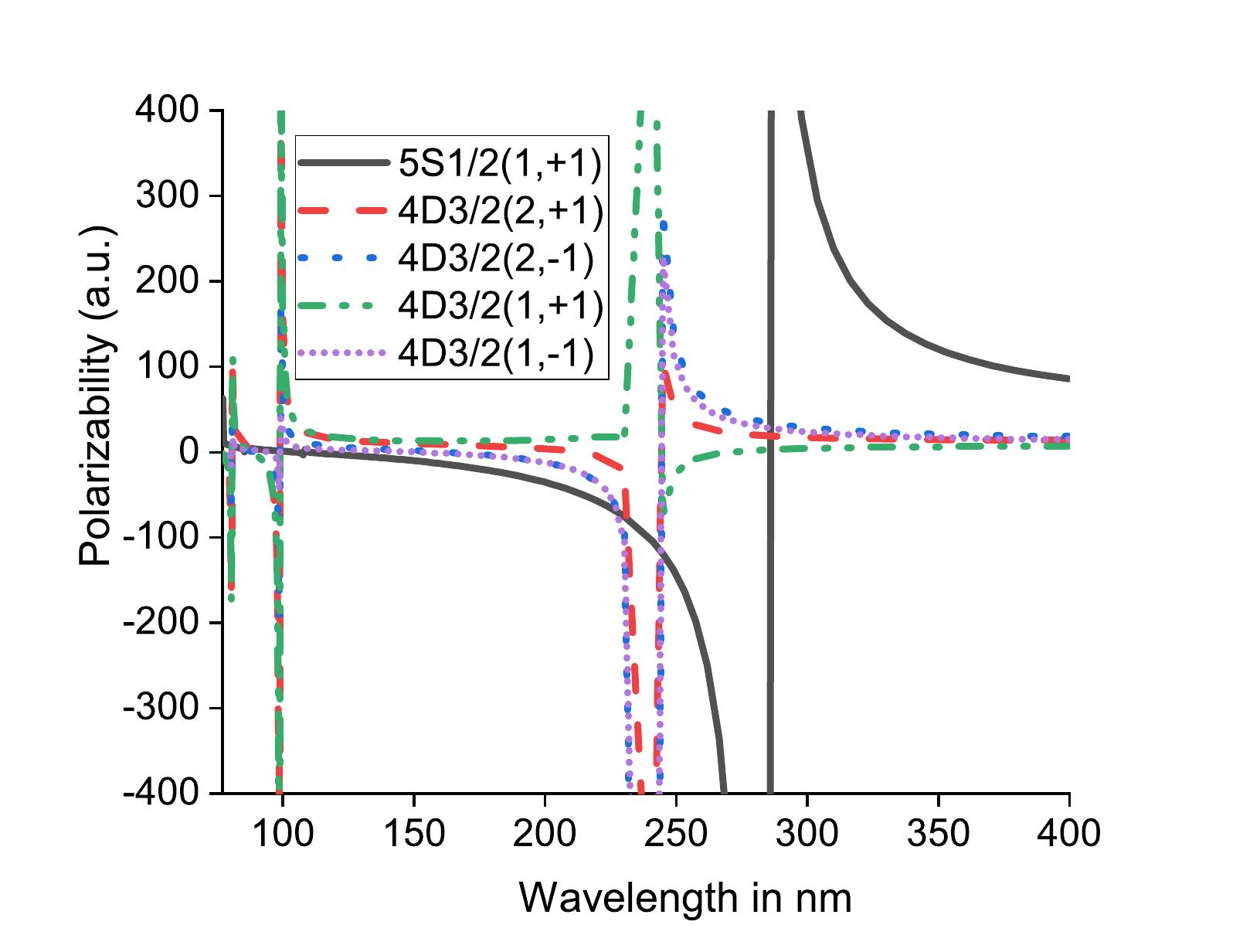}}
\subfloat[]{\includegraphics[width=0.52\linewidth]{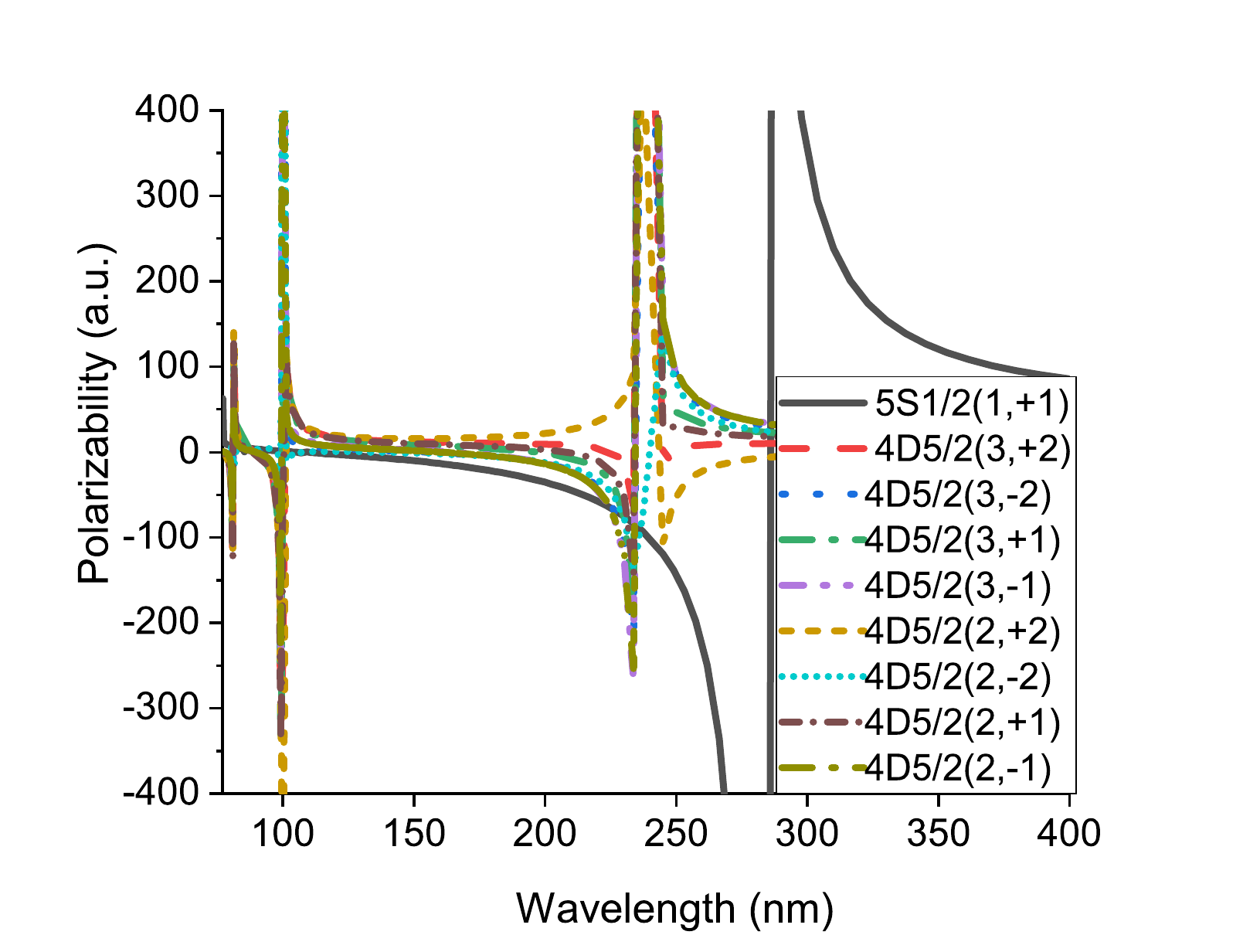}}
\caption{(Color online) Representation of Fig.~\ref{fig1}, but now by including the effect of hyperfine splittings in the fine-structure states 5$^2S_\frac{1}{2}$ and 4$^2D_{\frac{3}{2}, \frac{5}{2}}$.  (a) and (b) are for linearly polarized light. (c) and (d) are for circularly polarized light. Here the different $(F_v, M_{F_v})$ levels of the state n$^{2}L_{J_v}$ are indicated by n$LJ_{v}(F_v, M_{F_v})$ with n is the principal quantum number of the valence orbital and $L$ is the orbital angular momentum quantum number of the ion.}
\label{fig2}
\end{figure}

Fig.~\ref{fig2}. represents the variation  profiles of dynamic polarizabilities  for the hyperfine levels of the clock states within the same spectral region which is considered in Fig.~\ref{fig1}. For the circularly polarized light with $\sigma=+1$, the polarizability profiles of 4$^2D_{\frac{3}{2}}\left(F_v=2, M_{F_v}=\pm 2\right)$, 4$^2D_{\frac{5}{2}}\left(F_v=3, M_{F_v}=\pm 3\right)$ and 5$^2S_{\frac{1}{2}}\left(F_v=1, M_{F_v}=\pm 1\right)$ states are same as the profiles of 4$^2D_{\frac{3}{2}}\left(M_{J_v}=\pm \frac{3}{2}\right)$, 4$^2D_{\frac{5}{2}}\left(M_{J_v}=\pm \frac{5}{2}\right)$ and 5$^2S_{\frac{1}{2}}\left(M_{J_v}=\pm \frac{1}{2}\right)$ states, respectively. This is because of the unit value of the multiplication factors in Eq.~(\ref{8}) and Eq.~(\ref{9}) which relate $\alpha_{vF}^{(i)}(\omega)$ with $\alpha_{v}^{(i)}(\omega)$  for these states with $i=1, 2$. Also the polarizability values are same for 4$^2D_{\frac{3}{2}}\left(F_v=2, M_{F_v}=0\right)$ and 4$^2D_{\frac{3}{2}}\left(F_v=1, M_{F_v}=0\right)$ states. We have not shown the dynamic profile of the states separately in the figure for which profile is same.

In Table~\ref{IV}, we tabulate the magic wavelengths due to the transitions between the hyperfine levels of the clock states for a linearly and the circularly polarized light. Similar to Table~\ref{II}, we get few more magic wavelengths for the circularly polarized light compared to a linearly polarized light due to the presence of vector part of valence polarizability in the former. From a comparison between Table~\ref{IV} and Table~\ref{II}, it is obvious that the hyperfine splitting can induce small but noticeable changes both in the magic wavelengths and corresponding polarizabilities with more degrees of freedom in the choices of $(F_v,M_{F_v})$ combinations for each $(J_v,M_{J_v})$.

In order to calculate theoretical uncertainties in the RCCSD(T) hyperfine $A$ values in Table IV, we classify atomic states of Y$^{2+}$ in two different classes. One class (Class I) has the correlation correction by less than 50\%: 4$^2D_{\frac{3}{2}}$, 5$^2S_{\frac{1}{2}}$, 5$^2P_{\frac{1}{2}}$, 5$^2P_{\frac{3}{2}}$, 5$^2D_{\frac{3}{2}}$ and 5$^2D_{\frac{5}{2}}$, and the second class (Class II) has it by more than 50\%: 4$^2D_{\frac{5}{2}}$, 4$^2F_{\frac{5}{2}}$, and 4$^2F_{\frac{7}{2}}$. We believe the uncertainties in the hyperfine values are dominated by the uncertainties in the correlation corrections computed by our present RCCSD(T) method. Therefore, the uncertainties in the Class II states are supposed to be higher than the uncertainties in the Class I states. There can be small amount of uncertainties in the DF values as well, which we can approximately calculate by comparing the expectation values of $1/r$ for these states as computed by using our optimized GTO basis functions and as obtained by using the numerical DF wavefunctions of GRASP 92 code \cite{Parpia2006}. The $1/r$ values can certainly be one of the parameters to judge the accuracy of the DF wavefunctions near to the nuclear region where the hyperfine values are peaked. Nevertheless, as far as the correlation is concerned, the spread in the correlation corrections and therefore, in the total hyperfine values can be roughly estimated by comparing the present RCCSD(T) values with the corresponding values obtained from another similar method such as SDpT (linearized coupled-cluster with single, double and partial triple corrections) \cite{Safronova2010a, Safronova2010b} which can also provide very highly accurate results \cite{Dutta2014}. However, for Y$^{2+}$, no such SDpT hyperfine values are available to our knowledge. Therefore, we calculate the hyperfine values for Sr$^{+}$, which is also an element of Rb-isoelectronic sequence like Y$^{2+}$, by the present RCCSD(T) method and compare these values with the corresponding hyperfine values as calculated by the SDpT method of Safronova \cite{Safronova2010a}. Such a comparison is available for few states in Ref.\cite{Dutta2014}. Nevertheless, we find maximum discrepancy of 1.8\% in the comparison for all the states belong to the Class I. Our observations and experience of calculating hyperfine values for isoelectronic sequence of alkali-metal-like atoms \cite{Naren2013, Dutta2014, Narenthesis} say that for these states the discrepancy should not be more than $\pm$1.8\% for Y$^{2+}$, which is more ionized than Sr$^+$. This is mainly due to the fact that with increasing ionization in a particular isoelectronic sequence the correlation correction decreases \cite{Naren2013, Dutta2012}, and RCCSD(T) and SDpT or SD (SDpT without partial triple) values approach towards more and more close agreement usually \cite{Narenthesis}. Therefore, we can assume maximum uncertainty of $\pm$2\% for the hyperfine values of Class I states of Y$^{2+}$ due to discrepancy in correlation correction. However for Class II states, for which the correlation correction discrepancy between RCCSD(T) and SDpT or SD methods decreases from singly to doubly ionized systems along an isoelectronic sequence of an alkali-metal-atom \cite{Narenthesis}, a different strategy is considered. Here we compare the correlation correction in the hyperfine value of 4$^2D_{\frac{5}{2}}$ state as calculated by the present RCCSD(T) method and as calculated by the SDpT method of Safronova for Sr$^{+}$ \cite{Safronova2010a}. We find discrepancy of around 5\% in this correlation value. Now for Y$^{2+}$ which is more ionized than Sr$^{+}$, this discrepancy is expected to be lower than 5\%; even then, let us assume that this discrepancy is still 5\% for the purpose of calculating uncertainty. This 5\% discrepancy in correlation can invoke an uncertainty of around $\pm$7\% in the RCCSD(T) hyperfine  value of 4$^2D_{\frac{5}{2}}$ state of Y$^{2+}$. Now as for 4$^2F_{\frac{5}{2}}$ and 4$^2F_{\frac{7}{2}}$ states, no SDpT values are available in the work of Safronova for Sr$^{+}$, it is very difficult to compute the uncertainty values for these two states. Therefore, with a very rough assumption, if we assume the same amount of 5\% uncertainty in the correlation corrections of 4$^2F_{\frac{5}{2}}$ and 4$^2F_{\frac{7}{2}}$ states hold as well for Sr$^{+}$ ion, then one should not expect uncertainties by more than $\pm$7\% in the RCCSD(T) hyperfine values for these two states of Y$^{2+}$ also. Therefore, accounting the preciseness of DF wavefunctions near to the nuclear region, approximate contributions from the Breit interactions \cite{Dutta2012}, and from a comparative assessment between the RCCSD(T) and SDpT results for Sr$^{+}$, we estimate the theoretical uncertainties in the RCCSD(T) hyperfine $A$ values in Table~\ref{III} of around $\pm$3\% for the Class I states and $\pm$8\% for the Class II states.

To estimate uncertainty in the calculated magic wavelengths for Y$^{2+}$,  we reevaluate the magic wavelengths by replacing only our calculated RCCSD(T) dipole matrix elements by the corresponding available dipole matrix elements in the literature \cite{Safronova2013}  which were calculated by using an all-order RMBPT method. The transition matrix elements which we find in their work \cite{Safronova2013} are 5$^2S_{\frac{1}{2}}$-5$^2P_{\frac{1}{2}, \frac{3}{2}}$, 4$^2D_{\frac{3}{2}, \frac{5}{2}}$-$n$$^2P_{\frac{1}{2}, \frac{3}{2}}$ with $n =$ 5-7 and 4$^2D_{\frac{3}{2}, \frac{5}{2}}$-$m$$^2F_{\frac{5}{2}, \frac{7}{2}}$ with $m =$ 4-6. The maximum difference between these recalculated magic wavelengths and the corresponding magic wavelengths  presented in Table~\ref{II} and Table~\ref{IV} is considered as the uncertainty in the latter values which is estimated around $\pm$1\%.
  
\subsection{Tune-out wavelengths for 4$^2D_{\frac{3}{2},\frac{5}{2}}$  and 5$^2S_{\frac{1}{2}}$ states for a linearly polarized light}

We also report a few tune-out wavelengths in Table~\ref{V} for the hyperfine levels of the 5$^2S_{\frac{1}{2}}$ and 4$^2D_{\frac{3}{2},\frac{5}{2}}$  states of Y$^{2+}$. In this table, we consider a linearly polarized light only. There is an advantage of calculating tune-out wavelengths for a linearly polarized light. Table~\ref{V} displays the wavelength at which the total polarizability of a particular hyperfine state becomes zero for a linearly polarized light. However, due to the non-zero vector polarizability contribution, the total polarizability for a circularly polarized light does not become zero at the same wavelength for the same hyperfine state. As an example: the vector contribution to the total polarizability of the ground state $4^2D_{\frac{3}{2}}$ with $F_v=2, M_{F_v}=\pm2 $ at the  tune-out wavelengths 230.83 nm and 215.32 nm  become 414.14 a.u. and 25.84 a.u., respectively. Whereas, for the $F_v=1, M_{F_v}=\pm 1$ hyperfine component of the same fine-structure state, the vector polarizability values become 53.86 a.u. and 14.98 a.u. at the presented tune-out wavelengths 244.36 nm and 204.41 nm, respectively. These values can be used to calculate fictitious magnetic field induced by the circular polarization of light \cite{Albrecht2016}. The prior knowledge of the zero Stark-shift wavelengths for a particular atomic state can be advantageous for an accurate trap-insensitive experimental measurement \cite{ Leonard2015, Schmidt2016, Fallon2016}.

 \section{Conclusion}
In this present work, we have determined the magic wavelengths corresponding to the 5$^2S_{\frac{1}{2}}$-4$^2D_{\frac{3}{2},\frac{5}{2}}$ clock transitions of $^{89}$Y$^{2+}$ for two different types of polarization (linear and circular) of the projected light beam. The magic wavelengths  span between the vacuum ultraviolet to the near ultraviolet region of the electromagnetic spectrum. The data are presented both at the fine-structure and the associated hyperfine levels of the atomic states, and this gives an understanding of how much the hyperfine interaction can affect the magic trapping conditions viable for the fine-structure states. Indeed we have found slight modifications in the magic wavelength values after imposing hyperfine splitting on the fine-structure clock states. But most importantly, the number of magic wavelengths is increased. Many of these magic wavelengths can have potential applications for trap-related experimental explorations. Irrespective of the nature of the polarization of light, our calculations show that the polarizabilities at the magic wavelengths between 200 nm and 300 nm are higher, and thus, these wavelengths are more important for the experimental purposes. The calculated tune-out wavelengths in the present work have important applications to perform trap-insensitive experiments. Besides, we demonstrate quantitatively the advantage of using a circularly polarized light which yields extra trapping potential from the vector part of the polarizability.  The calculated static polarizabilities for 4$^2D_{\frac{3}{2},\frac{5}{2}}$, 5$^2S_{\frac{1}{2}}$ and 5$^2P_{\frac{1}{2}, \frac{3}{2}}$ states are useful to calculate the black-body radiation shifts of the transition frequencies among these states at a definite temperature. The hyperfine constants for most of the states of $^{89}$Y$^{2+}$ are reported for the first time in the literature to the best of our knowledge.

\begin{table}[htb]
	\centering
	\caption{Total static  polarizabilities $\alpha_{v}(0)$ are presented in a.u. along with contributions from scalar $\alpha_v^{(0)}(0)$,  core $\alpha^C_v(0)$, core-valence $\alpha^{VC}_v(0)$ and tensor $\alpha_v^{(2)}(0)$ parts. ``Other" refers to the corresponding values obtained from the work of Safronova and Safronova \cite{Safronova2013}. ``Our" refers to our calculated values which are rounded up to two decimal places.}
	\begin{tabular}{l c c c c c c c c c c c c c c}
		\hline\hline
		State & \multicolumn{3}{c}{$\alpha_v^{(0)}(0)$}& \multicolumn{3}{c}{$\alpha^C_v(0)$}& \multicolumn{3}{c}{$\alpha^{VC}_v(0)$}& \multicolumn{3}{c}{$\alpha_v^{(2)}(0)$}& \multicolumn{2}{c}{$\alpha_{v}(0)$}\\
		& Our  & Other && Our  & Other && Our  & Other && Our  & Other&& Our  & Other\\
		\hline
4$^2D_{\frac{3}{2}}$ & 6.89 & 6.742(26) && 4.28 & 4.048 && -0.33 & -0.313 && -3.48 & -3.45(2) && 7.36 & 7.03(4)\\
4$^2D_{\frac{5}{2}}$ & 6.93 & 6.815(32) && 4.28 & 4.048 && -0.36 & -0.341 && -4.86 & -4.81(3) && 5.99 & 5.71(4)\\
5$^2S_{\frac{1}{2}}$ & 42.07 & 40.64(17) && 4.28 & 4.048 && -0.17 & -0.17 && 0.00 & 0.00 && 46.18 & 44.5(2)\\
5$^2P_{\frac{1}{2}}$ & 8.75 & - && 4.28 & - && 0.00 & - && 0.00 & - && 13.03 &- \\
5$^2P_{\frac{3}{2}}$ & 12.02 & - && 4.28 & - && 0.00 & - && 5.61 & - && 21.91 &- \\
\hline

\end{tabular}
\label{table1}
\end{table}

\begin{table}[htb]
	\centering
	\caption{Contributions to $\alpha^V_{v}(0)$ (in a.u.) of the clock states from the intermediate states considered at the RCCSD(T), RMBPT(2), and DF levels.}
	\begin{tabular}{c r r r c r c }
		\hline\hline
State  && RCCSD(T) && RMBPT(2) && DF\\
		\hline
4$^2D_{\frac{3}{2}}$ &&  3.40233 &&  0.01070  && 0.00006\\
4$^2D_{\frac{5}{2}}$ &&  2.05729 &&  0.01265  && 0.00015\\		
5$^2S_{\frac{1}{2}}$ && 42.07169 &&  0.00089  && 0.00003\\ 
\hline
\end{tabular}
\label{table11}
\end{table}

\clearpage

\clearpage

\clearpage
\centering
\LTcapwidth=\linewidth	
\begin{longtable}{c r r r r r  r r r r r r r r r}
	
	\caption{Magic wavelengths $ \lambda_{\text{magic}}$ (in nm) with corresponding polarizabilities $\alpha_{\text{magic}}$ (in a.u.)  for 5$^2S_{\frac{1}{2}}$ to 4$^2D_{\frac{3}{2}}$ and 5$^2S_{\frac{1}{2}}$ to 4$^2D_{\frac{5}{2}}$ transitions due to linearly and circularly polarized ($\sigma$=+1) light. Notation $({J_v,M_{J_v}}) - ({J_v^{\prime},M_{J_v^{\prime}}})$ in the second row indicates the transition 5$^2S_{(J_v,M_{J_v})}$ $-$ 4$^2D_{(J_v^{\prime},M_{J_v^{\prime}})}$.}\\
	\hline\hline
	
	\multicolumn{3}{c}{Linearly Polarized} & \multicolumn{12}{c}{Circularly Polarized}\\
	\cline{1-3}   \cline{5-15}  
	\multicolumn{3}{c}{$(\frac{1}{2},\frac{1}{2})$ - $(J_v^{\prime},M_{J_v^{\prime}})$}  &  \multicolumn{3}{c}{$(\frac{1}{2}$,-$\frac{1}{2})$-$(J_v^{\prime}$,-$M_{J_v^{\prime}})$}& \multicolumn{3}{c}{$(\frac{1}{2}$,-$\frac{1}{2})$-$(J_v^{\prime}$,$M_{J_v^{\prime}})$}&
	\multicolumn{3}{c}{$(\frac{1}{2}$,$\frac{1}{2})$-$(J_v^{\prime}$,-$M_{J_v^{\prime}})$}& \multicolumn{3}{c}{$(\frac{1}{2}$,$\frac{1}{2})$-$(J_v^{\prime}$,$M_{J_v^{\prime}})$} \\
\cline{1-3}	\cline{5-6} \cline{8-9}\cline{11-12}\cline{14-15}
	$(J_v^{\prime},M_{J_v^{\prime}})$ &$\lambda_{\text{magic}}$ & $\alpha_{\text{magic}}$&&$\lambda_{\text{magic}}$ & $\alpha_{\text{magic}}$&&$ \lambda_{\text{magic}}$ & $\alpha_{\text{magic}}$&&$ \lambda_{\text{magic}}$ & $\alpha_{\text{magic}}$&&$ \lambda_{\text{magic}}$ & $\alpha_{\text{magic}}$ \\
	\hline
	\endfirsthead
	
	\multicolumn{15}{c}%
	{{\bfseries \tablename\ \thetable{} -- continued from previous page}} \\
	\hline \hline
	
	\hline 
		\multicolumn{3}{c}{Linearly Polarized} & \multicolumn{12}{c}{Circularly Polarized}\\
	\cline{1-3}   \cline{5-15}  
	\multicolumn{3}{c}{$(\frac{1}{2},\frac{1}{2})$ - $(J_v^{\prime},M_{J_v^{\prime}})$}  &  \multicolumn{3}{c}{$(\frac{1}{2}$,-$\frac{1}{2})$-$(J_v^{\prime},M_{J_v^{\prime}})$}& \multicolumn{3}{c}{$(\frac{1}{2}$,-$\frac{1}{2})$-$(J_v^{\prime}$,-$M_{J_v^{\prime}})$}&
	\multicolumn{3}{c}{$(\frac{1}{2}$,$\frac{1}{2})$-$(J_v^{\prime}$,$M_{J_v^{\prime}})$}& \multicolumn{3}{c}{$(\frac{1}{2}$,$\frac{1}{2})$-$(J_v^{\prime}$,-$M_{J_v^{\prime}})$} \\
\cline{1-3}	\cline{5-6} \cline{8-9}\cline{11-12}\cline{14-15}
	$(J_v^{\prime},M_{J_v^{\prime}})$ &$\lambda_{\text{magic}}$ & $\alpha_{\text{magic}}$&&$\lambda_{\text{magic}}$ & $\alpha_{\text{magic}}$&&$ \lambda_{\text{magic}}$ & $\alpha_{\text{magic}}$&&$ \lambda_{\text{magic}}$ & $\alpha_{\text{magic}}$&&$ \lambda_{\text{magic}}$ & $\alpha_{\text{magic}}$ \\
	
	\hline 
	\endhead
	
	\hline \multicolumn{3}{r}{{Continued on next page}} \\ 
	\endfoot
	
	\hline \hline
	\endlastfoot
	
	($\frac{3}{2}, \frac{3}{2}$)& 293.77 & 9.57&  	&	294.83	& 22.16 &	 	& 294.77	& -3.11 & 	&	286.02	&	24.91	&	 	&	286.00	&	-5.11 \\
	&-&-&& 244.18	&	-97.23	&	&	247.36	&	-107.53 &&	244.12	&	-116.35	&		&	246.54	&	-126.90 \\
	&-&-&& 226.14	&	-59.47	&	&	244.79	&	-98.92 &&	232.54	&	-82.32	&		&	244.85	&	-119.02\\
	&-&-&&	101.08	&	0.59	&	&-	&- &&	101.09	&	0.68	&	&-	&-	\\
	& 98.96 & 1.10 &&	98.96	&	1.06	&		&	98.96	&	1.06 &&	98.96	&	1.15	&	& 98.96	& 1.15 \\
	& 85.39 & 4.52 &&	85.31   &	3.35 &		&	85.35	&	5.79 &&	85.50&	3.36&		&	85.61	&	5.22\\
	& 80.63 & 6.59 & &	80.52	&	6.45	&		&	80.63	&	6.36 &&	80.50	&	6.93	&		&	80.08	&	7.32\\
	($\frac{3}{2}, \frac{1}{2}$)& 293.84 &28.79&  	&	294.85	&	32.99	&		&	294.83	&	24.58 &	&	286.02	&	36.85	&		&	286.02	&	26.85\\
	&244.26 & -107.13 & &	244.32	&	-97.61	&		&	244.20	&	-97.30&&	244.29	&	-116.94	&		&	244.16	&	-116.45\\
	&233.83 &-78.13 &	&	226.68	&	-60.24	&		&	230.62	&	-67.61 &	&227.31	&	-70.05	&		&	231.99	&	-81.06\\
	& 98.96& 1.10& &	98.96	&	1.06	&	 &	98.96	&	1.06 &&	98.96	&	1.14	&		&	98.96	&	1.14\\
	& 85.38 & 3.88& &	85.73	&	3.07	&		&	85.80	&	3.48 &&	85.16	&	3.81	&		&	85.44	&	4.23\\
	& 80.63 & 6.59& &	80.63	&	6.36	&		&	80.63	&	6.36 &&	80.63	&	6.82	&		&	80.63	&	6.83\\
	($\frac{5}{2}, \frac{5}{2}$)& 293.75 & 6.26 &		&	294.83	&	19.71	&		&	294.77	&	-7.21	&		&	286.02	&	22.18	&		&	286.00	&	-9.64\\
	&-&-&		&	238.33	&	-81.99	&		&	246.62	&	-105.03	&	&	238.26	&	-96.02	&	&	245.47	&	-121.85\\
	&-&-&		&	229.28	&	-64.40	&		&	239.15	&	-83.97	&		&	230.70	&	-76.69	&		&	239.23	&	-99.04\\
	&-&-&		&	113.16	&	-0.82	&		&-	&-	&		&	112.17	&	-0.95	&&-	&-	\\
	&-&-&		&	100.53	&	0.72	&		&	100.40	&	0.75	&		&	100.53	&	0.81	&		&	100.40	&	0.83\\
	& 99.54 & 0.98 &		&	99.30	&	0.99	&		&-	&-		&		&	99.30	&	1.07	&&-		&-	 \\
	& 85.40 & 5.04&		&	98.63	&	1.13	&		&	86.15	&	5.57	&		&	98.63	&	1.22	&		&	86.22	&	5.42 \\
	& 81.02 & 6.28 &		&	88.78	&	3.67	&		&	80.99	&	6.08	&		&	88.86	&	3.69	&		&	80.99	&	6.53\\
	
	($\frac{5}{2}, \frac{3}{2}$)& 293.82 & 21.05 &		&	294.85	&	28.94	&		&	294.81	&	12.81	&		&	286.02	&	32.16	&		&	286.00	&	13.07\\
	&233.91 & -78.33&		&	234.17	&	-73.14	&		&	233.72	&	-72.19	&		&	234.14	&	-84.68	&		&	233.69	&	-83.47\\
	& 229.71 & -69.82&		&	222.72	&	-54.69	&		&	232.28	&	-69.59	&		&	224.60	&	-64.94	&		&	232.91	&	-81.69\\
	&99.54 &  0.98 &		&	99.38	&	0.97	&		&	100.78	&	0.66	&		&	99.38	&	1.06	&		&	100.25	&	0.87\\
	&85.40 & 4.65&		&	85.73	&	3.08	&		&	85.99	&	4.64	&		&	85.04	&	3.62	&		&	85.89	&	4.92\\
	&81.02 & 6.28&		&81.08	&	6.01	&		&	81.00	&	6.07	&		&	81.08	&	6.46	&		&	81.00	&	6.52
	\\
	($\frac{5}{2}, \frac{1}{2}$)& 293.87 & 28.44&		&	294.85	&	30.87	&		&	294.83	&	25.50	&		&	286.02	&	33.97	&		&	286.02	&	27.61\\
	& 233.99 & -78.51 &		&	234.05	&	-72.89	&		&	233.95	&	-72.67	&		&	234.03	&	-84.39	&		&	233.93	&	-84.11\\
	& 225.48 & -62.52&		&	222.50	&	-54.40	&		&	226.83	&	-60.48	&		&	224.26	&	-64.35	&		&	227.71	&	-70.62\\
	& 99.54  & 0.98 &		&	99.47	&	0.95	&		&	99.66	&	0.91	&		&99.47	&	1.03	&		&	99.66	&	1.00\\
	& 85.39 & 4.46&		&	85.79	&	3.43	&		&	85.88	&	3.94	&		&	85.37	&	4.13	&		&	85.62	&	4.52\\
	& 81.02 & 6.28&		&	81.03	&	6.05	&		&	81.02	&	6.06	&		&	81.03	&	6.50	&		&	81.02	&	6.51
	\label{II}
\end{longtable}

\begin{table} [h!]
	\centering
	\caption{Hyperfine $A$ constants and hyperfine splitting in the unit of MHz.}
	\begin{tabular}{c r r r r r r r}
		\hline\hline
		State& \multicolumn{4}{c}{Hyperfine \textit{A} constant} & $F_1\rightarrow F_2$& \multicolumn{2}{c}{Hyperfine splitting}\\ 
		\cline{2-5}\cline{7-8}
		& DF & RCCSD(T)& \multicolumn{2}{c}{Other \cite{Vormawah2018}} & & RCCSD(T)& Other \cite{Crawford1949}  \\ 
		\cline{4-5}
		&&&MCDF& Exp.&&&\\
		\hline
		4$^2D_{\frac{3}{2}}$&	79.55& 102.74&-&-&2$\rightarrow$1&	205.47&-\\
		4$^2D_{\frac{5}{2}}$&	33.20& 14.25&-&-&3$\rightarrow$2&	42.74&-\\
		4$^2F_{\frac{5}{2}}$&	 0.48& 0.86	&-&-&3$\rightarrow$2&	2.59&-\\
		4$^2F_{\frac{7}{2}}$&	-0.27& -0.88&-&-&4$\rightarrow$3&	3.52&-\\
		5$^2S_{\frac{1}{2}}$&	1441.92& 1793.19&1780&1803(5) &1$\rightarrow$0&1793.19&1920$\pm$150\\
		5$^2P_{\frac{1}{2}}$&	284.44& 371.47&352&391(5) &1$\rightarrow$0&	371.47&-\\
		5$^2P_{\frac{3}{2}}$&	49.57& 73.86&-&-&2$\rightarrow$1&	147.72&128\\
		5$^2D_{\frac{3}{2}}$&	17.12& 21.63&-&-&2$\rightarrow$1&   43.25&-\\
		5$^2D_{\frac{5}{2}}$&	 7.19& 8.10	&-&-&3$\rightarrow$2&   24.29&-\\
		\hline
	\end{tabular}
	\label{III}
\end{table}
\clearpage

\begin{center}
	
	\begin{longtable}{c r r r r r r r r r r r r r r}
		\caption{Magic wavelengths $\lambda_{\text{magic}}$ (in nm) with corresponding total polarizabilities $\alpha_{\text{magic}}$ (in a.u.) for the 5$^2S_\frac{1}{2}$ $-$ 4$^2D_{\frac{3}{2}, \frac{5}{2}}$ transitions for linearly and circularly polarized ($\sigma$=+1) lights. Notation $(F_v,M_{F_v})-(F_v^{\prime},M_{F_v^{\prime}})$ in the second row indicates the transition 5$^2S_{(J_v,M_{J_v})}(F_v,M_{F_v})$ $-$  4$^2D_{(J_v^{\prime},M_{J_v^{\prime}})}(F_v^{\prime},M_{F_v^{\prime}})$.}\\
		\hline
	\multicolumn{3}{c}{Linearly Polarized} & \multicolumn{12}{c}{Circularly Polarized}\\
	\cline{1-3}   \cline{5-15}  
	\multicolumn{3}{c}{$(1,1)$-$(F_v^{\prime},M_{F_v^{\prime}})$}  &  \multicolumn{3}{c}{$(1$,-$1)$-$(F_v^{\prime}$,-$M_{F_v^{\prime}})$}& \multicolumn{3}{c}{$(1$,-$1)$-$(F_v^{\prime}$,$M_{F_v^{\prime}})$} &
	\multicolumn{3}{c}{$(1,1)$-$(F_v^{\prime}$,-$M_{F_v^{\prime}})$}& \multicolumn{3}{c}{$(1$, $1)$-$(F_v^{\prime}$,$M_{F_v^{\prime}})$} \\
\cline{1-3}	\cline{5-6} \cline{8-9}\cline{11-12}\cline{14-15}
	[$J_v^{\prime}$]$(F_v^{\prime},M_{F_v^{\prime}})$ &$\lambda_{\text{magic}}$ & $\alpha_{\text{magic}}$&&$\lambda_{\text{magic}}$ & $\alpha_{\text{magic}}$&&$ \lambda_{\text{magic}}$ & $\alpha_{\text{magic}}$&&$ \lambda_{\text{magic}}$ & $\alpha_{\text{magic}}$&&$ \lambda_{\text{magic}}$ & $\alpha_{\text{magic}}$ \\		
		
		\hline
		\endfirsthead
		
		\multicolumn{15}{c}%
		{{\bfseries \tablename\ \thetable{} -- continued from previous page}} \\
		\hline \hline
		
		\hline 
		
	\multicolumn{3}{c}{Linearly Polarized} & \multicolumn{12}{c}{Circularly Polarized}\\
	\cline{1-3}   \cline{5-15}  
	\multicolumn{3}{c}{$(1,1)$-$(F_v^{\prime},M_{F_v^{\prime}})$}  &  \multicolumn{3}{c}{$(1$,-$1)$-$(F_v^{\prime},M_{F_v^{\prime}})$}& \multicolumn{3}{c}{$(1$,-$1)$-$(F_v^{\prime}$,-$M_{F_v^{\prime}})$} &
	\multicolumn{3}{c}{$(1,1)$-$(F_v^{\prime},M_{F_v^{\prime}})$}& \multicolumn{3}{c}{$(1$, $1)$-$(F_v^{\prime}$,-$M_{F_v^{\prime}})$} \\
\cline{1-3}	\cline{5-6} \cline{8-9}\cline{11-12}\cline{14-15}
	[$J_v^{\prime}$]$(F_v^{\prime},M_{F_v^{\prime}})$ &$\lambda_{\text{magic}}$ & $\alpha_{\text{magic}}$&&$\lambda_{\text{magic}}$ & $\alpha_{\text{magic}}$&&$ \lambda_{\text{magic}}$ & $\alpha_{\text{magic}}$&&$ \lambda_{\text{magic}}$ & $\alpha_{\text{magic}}$&&$ \lambda_{\text{magic}}$ & $\alpha_{\text{magic}}$ \\

		\hline 
		\endhead
		
		\hline \multicolumn{3}{r}{{Continued on next page}} \\ 
		\endfoot
		
		\hline \hline
		\endlastfoot

		[$\frac{3}{2}$] (2,2)& 293.77 & 9.57&  	&	294.83	& 22.16 &	 	& 294.77	& -3.11 & 	&	286.02	&	24.91	&	 	&	286.00	&	-5.11 \\
		&-&-&& 244.18	&	-97.23	&	&	247.36	&	-107.53 &&	244.12	&	-116.35	&		&	246.54	&	-126.90 \\
		&-&-&& 226.14	&	-59.47	&	&	244.79	&	-98.92 &&	232.54	&	-82.32	&		&	244.85	&	-119.02\\
		&-&-&&	101.08	&	0.59	&	&-	&- &&	101.09	&	0.68	&	&-	&-	\\
		& 98.96 & 1.10 &&	98.96	&	1.06	&		&	98.96	&	1.06 &&	98.96	&	1.15	&	& 98.96	& 1.15 \\
		& 85.39 & 4.52 &&	85.31   &	3.35 &		&	85.35	&	5.79 &&	85.50&	3.36&		&	85.61	&	5.22\\
		& 80.63 & 6.59 & &	80.52	&	6.45	&		&	80.63	&	6.36 &&	80.50	&	6.93	&		&	80.08	&	7.32\\
		$[\frac{3}{2}]$, (2,1) & 293.85 & 24.2 & & 294.88	&	30.28& & 294.85	& 17.66& & 286.03	&	32.76& & 286.00 &	18.86\\
		&244.19&-106.87&&244.30	&	-97.55& &243.93	&	-96.51&& 244.27& -116.84&& 243.82&-115\\
		& 231.91 & -74.45&& 226.53	&	-60.02&&227.81	&	-62.45&&226.97&	-69.28&&237.54&-93.76\\
		& 98.96 & 1.10 &&98.97	&	1.06&& 98.96	&	1.06&& 99.41	&	1.05&&98.96	& 1.14\\
		&85.38&4.06&&85.30	&	3.53 && 85.34	&	4.71&& 85.51	&	3.44&&85.56	&	4.28\\
		&80.63&6.59&&80.63	&	6.36&&80.63	&	6.36 && 81.06	&	6.48&& 80.63	&	6.83\\
		$[\frac{3}{2}]$, (2/1,0)&293.87&29.07&&-&&-&-&-&&-&-&&-&-\\
		&244.27&-107.13&&-&&-&-&-&&-&-&&-&-\\
		&230.62& -71.63&&-&&-&-&-&&-&-&&-&-\\
		&98.96& 1.10&&-&&-&-&-&&-&-&&-&-\\
		&85.36& 3.92&&-&&-&-&-&&-&-&&-&-\\
		&80.63& 6.59&&-&&-&-&-&&-&-&&-&-\\
		$[\frac{3}{2}]$, (1,1)&293.81	&	14.45& & 294.86	&	24.87&  & 294.82	&	3.82&  & 286.03&	27.89&  &286.02 &	2.88\\
		&243.55	&	-104.76&&244.24	&	-97.36&&-&-&&244.20	&	-116.56&&-&-\\
		&237.45	&	-86.66&&226.26	&	-59.63&&-&-&&226.74	&	-68.71&&-&-\\
		&98.96	&	1.10&&98.97	&	1.06&& 98.96	&	1.06&&98.97	&	1.14&&98.96	&	1.15\\
		&85.39	&	4.36&&85.34	&	3.40&&85.37	&	5.38&& 85.50	&	3.36&& 85.60	&	4.91\\
		&80.63	&	6.59&&84.20	&	3.78&&-&-&&84.66	&	3.61&& -&-\\
		&-&-&&82.05	&	5.34&& -&-&& 81.71	&	6.00&& -&-\\
		&-&-&&80.66	&	6.34&& 80.63	&	6.36&& 80.66	&	6.80&& 80.63	&	6.82\\
		$[\frac{5}{2}]$, (3,3)& 293.75 & 6.26 &		&	294.83	&	19.71	&		&	294.77	&	-7.21	&		&	286.02	&	22.18	&		&	286.00	&	-9.64\\
		&-&-&		&	238.33	&	-81.99	&		&	246.62	&	-105.03	&	&	238.26	&	-96.02	&	&	245.47	&	-121.85\\
		&-&-&		&	229.28	&	-64.40	&		&	239.15	&	-83.97	&		&	230.70	&	-76.69	&		&	239.23	&	-99.04\\
		&-&-&		&	113.16	&	-0.82	&		&-	&-	&		&	112.17	&	-0.95	&&-	&-	\\
		&-&-&		&	100.53	&	0.72	&		&	100.40	&	0.75	&		&	100.53	&	0.81	&		&	100.40	&	0.83\\
		& 99.54 & 0.98 &		&	99.30	&	0.99	&		&-	&-		&		&	99.30	&	1.07	&&-		&-	 \\
		& 85.40 & 5.04&		&	98.63	&	1.13	&		&	86.15	&	5.57	&		&	98.63	&	1.22	&		&	86.22	&	5.42 \\
		& 81.02 & 6.28 &		&	88.78	&	3.67	&		&	80.99	&	6.08	&		&	88.86	&	3.69	&		&	80.99	&	6.53\\
		$[\frac{5}{2}]$, (3,2)&293.81 & 18.59 &   &294.83	&	26.81&  & 294.81	&	10.07 &  &286.02	&	29.79 &   & 286.00	&	9.99\\
		&233.86&-78.21&& 234.20	&	-73.18&& 243.21	&	-94.54&& 234.15	&	-84.70&& 242.84	& -111.49\\
		&230.63&-71.59&& 224.20	&	-56.72&& 240.77	&	-87.79&& 226.17	&	-67.66 &&240.82&-103.99\\
		&98.89&0.98&& 99.37	&	0.97&& 99.62&0.92&&98.72&1.06&&99.62&0.86\\
		&84.92&4.71&& 85.35	&	3.37 && 85.37&5.06&&85.02&3.28 &&85.15&5.55\\
		&80.59 &6.28 &&81.09	&	6.00&& 80.57	&	6.41 && 80.67&	6.45 && 80.57	&	6.52 \\
		$[\frac{5}{2}]$, (3,1)&293.86&25.97& & 294.87	&	29.59&  & 294.85	&	17.04& & 286.03	&	33.37& & 286.02	&	22.76\\
		&233.98&-78.47&& 234.13	&	-73.03&& 233.82	&	-72.39&& 234.07	&	-84.46&& 233.87	&	-83.94\\
		&227.00&-64.97&&222.65	&	-54.59&& 230.73	&	-66.81&& 224.38	&	-64.53&& 230.15	&	-75.42\\
		&99.54&0.98&&99.41	&	0.96&& 100.09	&	0.81&& 99.44	&	1.04&& 99.80	&	0.97\\
		&85.39&4.54&& 85.36	&	3.59&& 85.38	&	5.66&& 85.52	&	3.63&& 85.59	&	4.83\\
		&81.02&6.28&&81.06	&6.03&& 81.01	&	6.07&& 81.04	&	6.49&& 
		81.01	&	6.52\\
		$[\frac{5}{2}]$, (3/2,0)&293.87&27.06&&-&&-&-&&-&-&&-&-&-\\
		&230.46&-71.22&&-&&-&-&&-&-&&-&-&-\\
		&221.19&-56.04&&-&&-&-&&-&-&&-&-&-\\
		&98.90&1.12&&-&&-&-&&-&-&&-&-&-\\
		&85.38&3.62&&-&&-&-&&-&-&&-&-&-\\
		&80.59&6.62&&-&&-&-&&-&-&&-&-&-\\
		$[\frac{5}{2}]$, (2,2)&293.79 & 8.72 &  & 294.86	&	21.25 & & 294.81	&	-3.87&   & 286.02	&	23.85 &  &  286.01	&	-5.86\\
		&-&-&&237.43	&	-79.84 && 245.50	&	-101.19&&237.35	&	-93.14&& 244.83	&	-118.97\\
		&-&-&&227.89	&	-62.16&& 239.63	&	-85.05&& 229.69	&	-74.52&& 239.69	&	-100.44\\
		&-&-&&110.77	&	0.24&& -& -&& 109.63	&	0.10&&-&-\\
		&-&-&&106.08	&	-0.89&&-& -&&106.28	&	-0.80&&-&-\\
		&99.54	&	0.98&&97.78	&	1.32&&100.39	&	0.75&&97.76	&	1.40&&99.99	&	0.92\\
		&85.40	&	4.97&& 89.74	&	3.32&&85.39	&	7.05&& 89.93	&	3.33&& 85.68	&	6.37\\
		&81.02	&	6.28&& 85.33	&	2.91&& 80.99	&	6.08&& 85.48	&	2.93&& 80.99	&	6.53 \\
		$[\frac{5}{2}]$, (2,1)&293.85&22.44&  & 294.87	&	29.59&  & 294.85	&	17.04&   &286.03	&	32.76&  & 286.02	&	17.92\\
		&230.42&-71.14&&234.13	&	-73.03&& 233.82	&	-72.39&& 234.10	&	-84.55& & 233.79	&	-83.73\\
		&224.08&-60.28&&222.65	&	-54.59&&230.73	&	-66.81&& 224.49	&	-64.73&& 231.16	&	-77.72\\
		&98.90&1.12&& 98.76	&	0.96&& 99.61	&	0.81&& 98.76	&	1.05&& 99.61	&	0.90\\
		&85.38&3.83&& 84.85	&	3.62&& 84.87	&	5.73&& 85.03	&	3.44&& 85.13	&	5.13\\
		&80.59&6.62&& 80.63	&	6.03 && 80.58	&	6.07&& 80.63	&	6.48&& 80.58	&	6.52
		\label{IV}
	\end{longtable}
\end{center}

\clearpage
\begin{table} [h!]
	\centering
	\caption{Tune-out wavelengths (in nm) for Y$^{2+}$ at the hyperfine levels ($F_v$,$M_{F_v}$).}
	\begin{tabular}{r r r r r c r r r r r r}
		\hline\hline
		5$^2S_{\frac{1}{2}}$ & \multicolumn{4}{c}{4$^2D_{\frac{3}{2}}$}
		&& \multicolumn{6}{c}{4$^2D_{\frac{5}{2}}$} \\
		\cline{2-5}\cline{7-12}
		(1,1) & (2,2) & (2,1)& (2 or 1, 0)& (1,1) &&(3,3) & (3,2) & (3,1) & (3 or 2,0) & (2,2)& (2,1) \\
		\hline
		293.73 & 230.83 & 244.46 & 244.47 & 244.36 &&99.54 & 234.14 & 234.15 & 230.59 & 230.50 & 230.59 \\
		108.68 & 215.32 &186.85 & 180.59 & 204.41 && 91.04 & 193.06 & 180.46 & 175.24 & 220.54 & 181.86 \\
		& 98.96 & 98.96 & 98.96 & 98.96 && & 99.54 & 99.54 & 98.90 & 98.89 & 98.97 \\
		& 90.01 & 88.02 & 88.02 & 89.38 && & 89.19 & 88.45 & 87.74 & 90.06 & 88.16 \\
		& 80.61 & 80.62 & 80.62 & 80.62 && 81.00 &81.00 & 81.01 & 80.58 & 80.57 & 80.58 \\
		\hline
	\end{tabular}
	\label{V}
\end{table}

\end{document}